\documentclass[nofootinbib,notitlepage,superscriptaddress,10pt,aps,pra]{revtex4-1}

\usepackage[utf8x]{inputenc}
\usepackage{amsmath}
\usepackage{amssymb}
\usepackage{graphicx}
\usepackage[normalem]{ulem}
\usepackage{epsf,epsfig}
\usepackage{psfrag}
\usepackage{color}
\usepackage{multirow}
\usepackage{diagbox}
\usepackage{dsfont}
\usepackage{cancel}

\makeatletter
\newcommand\stroke[1]{\mathpalette\stroke@aux{#1}}
\def\stroke@aux#1#2{%
 \ooalign{%
  \hfil$#1^{\;\, \_\hspace{-0.05cm}\_}$\hfil\cr
  \hfil$#1#2$\hfil\cr
 }%
}
\makeatother

\begin{document}

\title{Mixing coproducts for theories with particle-dependent relativistic properties}

\author{Giovanni Amelino-Camelia}
\affiliation{indirizzo federico II}
\affiliation{INFN, sezionenapoli}

\author{Michelangelo Palmisano}
\affiliation{Dipartimento di Fisica, Universit\`a di Roma ``La Sapienza", P.le A. Moro 2, 00185 Roma, Italy}
\affiliation{INFN, Sez.~Roma1, P.le A. Moro 2, 00185 Roma, Italy}

\author{Michele Ronco}
\affiliation{Laboratoire de Physiqe Nucléaire et de Hautes Energies (LPNHE) UPMC, Case courrier 200, 4 place Jussieu, F-75005 Paris, France}
\affiliation{UPMC, Case courrier 200, 4 place Jussieu, F-75005 Paris, France}

\author{Giacomo D'Amico}
\affiliation{Dipartimento di Fisica, Universit\`a di Roma ``La Sapienza", P.le A. Moro 2, 00185 Roma, Italy}
\affiliation{INFN, Sez.~Roma1, P.le A. Moro 2, 00185 Roma, Italy}

\begin{abstract}
We analyze a few illustrative examples of scenarios in which relativistic symmetries are deformed by Planck-scale effects 
in particle-type-dependent manner. The novel mathematical structures required by such scenarios
are the mixing coproducts, which govern the (deformed) law of conservation of energy and momentum when particles
with different relativistic properties interact. We also comment on the relevance of these findings for recent
proposals concerning the possibility that neutrinos might have relativistic properties which are different from those of photons
and/or the possibility that composite particles might have relativistic properties which are different from those of fundamental ones.
\end{abstract}

\maketitle

\section{INTRODUCTION}

Over the last decade there have been several studies investigating the fate of relativistic symmetries at the Planck scale (see, {\it e.g.}, Refs.\cite{amelino2013quantum,magueijo2003generalized,amelino2012fate}). Particular interest has been devoted to scenarios such that the (inverse of the) Planck scale would set the minimum allowed value for wavelengths, and notably it has emerged that such a feature could be implemented within a fully relativistic picture \cite{amelino2001testable,amelino2011principle}. This is the realm of the so-called doubly-special (or deformed-special) relativistic (DSR) theories \cite{amelinoDSRijmpd1135,amelino2001testable,amelino2010doubly,kowalski2003non} where the Planck scale plays the role of a second relativistic invariant, in addition to the speed-of-light scale. Evidently these DSR scenarios require the adoption of deformed Poincar\'e transformations connecting inertial frames \cite{amelinoDSRijmpd1135}, and the associated new invariant laws may include indeed an oberver-independent minimum-wavelength law and/or an observer-independent modification of the dispersion/on-shellness relation (MDR). Another major implication of the DSR deformations of Poincar\'e transformations is that the laws of composition of momenta are no longer linear: in order to preserve their relativistic covariance they must be deformed matching the deformation of the Poincar\'e transformations.
 The mathematical formalism of Hopf algebras has been found to be a natural possibility\footnote{Note however that examples of DSR-relativistic scenarios not
 involving Hopf algebras have been provided in Refs.\cite{freidelDESITTERSNYDER,balena}.} for formalizing DSR-relativistic scenarios \cite{agostini2004hopf}, since Hopf algebras can be deformations of standard Lie algebras in the form of non-linear deformations of both commutation relations among symmetry generators and the so-called coproducts (on which the laws of composition of momenta are based; see later).

 So far all of these studies (with the only exception of the exploratory analysis in Ref.\cite{gacmixingold})
 focused on \textit{universal} deformations of the special-relativistic symmetries, {\it i.e.} deformations that affect identically all particle types.
Here we explore the possibility that the description of Planck-scale physics might require an additional level of complexity, namely different particles' kinematics might be dictated by different relativistic symmetries, {\it i.e.} we wonder whether, within the DSR framework, it is possible to consistently formulate relativistic theories that attribute different laws of kinematics to different particles, \textit{non-universal} deformations of relativistic symmetries. For the case of broken relativistic symmetries (preferred-frame scenarios) particle-dependent effects where first considered by Coleman and Glashow in Ref.\cite{coleman1997cosmic} and there were several developments of that research direction (see, {\it e.g.}, Ref.\cite{smeREVIEW}), but the possibility of particle-dependent properties within a relativistic picture has been so far only explored preliminarily by one of us in Ref.\cite{gacmixingold}, providing most results only
 at leading order in the deformation scale. We shall here report results valid to all orders in the deformation scale and consider a rather wide class of possible properties attributed to different types of particles. We also intend to show that these scenarios
 can be built using rather mild modifications of standard Hopf-algebra techniques. As we shall discuss extensively in the following sections, the key building block for our scenarios is a novel mathematical tool which we call \textit{mixing coproduct}, a generalization of
 the standard Hopf-algebra notion of coproduct. \\
 We shall here mostly postpone the analysis of the phenomenological implications of our
\textit{non-universal} deformations of relativistic symmetries, but the careful reader will notice how that ultimate objective guides
our technical efforts. In particular, while \textit{non-universal} deformations of relativistic symmetries
do not necessarily require modifications of the on-shell relation, in all of our case studies there is at least
one type of particle governed by modified on-shellness.
Indeed, one of the main reasons of interest in DSR-relativistic theories has been their rich phenomenology
when they involve modified on-shellness, a rare case of Planck-scale effect testable with
presently-available technologies \cite{grbgac1998,gacSMOLINprd2009,amelino2015icecube,jacob2007neutrinos}.
Of particular interest for the analysis we here report is that fact that some recent tests of modifications of on-shellness
for neutrinos have led to preliminarily encouraging results \cite{amelino2015icecube,Amelino-Camelia:2016ohi,amelino2016icecube,paperbyMA},
for values of the symmetry-deformation scale of about a tenth of the Planck scale, whereas for photons
other analyses have led to apparently robust bounds on the symmetry-deformation scale going all the way up to
the order of the Planck scale. In spite of the very preliminary nature of the mentioned neutrino studies, one cannot
avoid to wonder what would have to be the relativistic picture if it was actually true that modifications
of relativistic symmetries are stronger for neutrinos than for photons, leading us indeed to speculate
about non-universality of the deformation scheme \cite{abdo2009limit,aharonian2008limits}.

In addition to the present preliminary assessment of data on dispersion for photons and neutrinos further motivation
for non-universality is found upon contemplating macroscopic bodies (like soccerballs and planets) in a DSR picture:
it is easy to see \cite{amelino2011relative} that a phenomenologically viable DSR picture should ensure that the deformation
of relativistic properties fades away for macroscopic bodies, and there are known relativistic mechanisms
to enforce this property (see, {\it e.g.}, \cite{amelino2011relative} and references therein);
however it is still unclear what should be the DSR-relativistic kinematics applicable to processes involving a
fundamental particle and a more macroscopic systems. Assuming that indeed the DSR-deformation
of relativistic properties is, for example, stronger for the electron than for the bucky ball, what are the conservation laws
that one should apply for collisions between an electron and a bucky ball? 

Our paper is organized as follows. In Section II we briefly review a well-known example of \textit{universal} deformation of the Poincar\'e symmetries, based on the $\kappa$-Poincar\'e Hopf algebra.
In Section III we start showing how it is possible to generalize such a setup in order to allow for models where
different particles obey different symmetry laws.
The first case we consider is that of having two (or more) deformed Poincar\'e algebras with different
deformation scales, e.g. $\ell$ and $\ell'$. This gives us the possibility to introduce
the notion of \textit{mixing coproduct} for the formalization of the kinematics of such \textit{non-universal} deformed-symmetry models. In Section IV we show that a scenario previously advocated by Magueijo and Smolin \cite{magueijo2003generalized} can be equivalently
reformulated in terms of a specific mixing coproduct. In Section V we add a further element of complexity and explore the possibility to define a mixing coproduct between different algebras, focusing on the case of mixing coproducts involving the $\kappa$-Poincar\'e algebra and the standard Poincar\'e Lie algebra. In our closing Section VI we offer a perspective on our results and
some observations on possible future developments.

\section{Universal coproduct: $\kappa$-Poincar\'{e} case of study}

In preparation for discussing DSR scenarios with nonuniversal relativistic properties, we find
useful to briefly review the most studied DSR scenarios with universal relativistic properties, which is based on mathematical structures
found in the $\kappa$- Poincar\'e Hopf algebra (specifically the so-called \textit{bi-cross-product basis}
of the $\kappa$- Poincar\'e Hopf algebra). When written in the \textit{bi-cross-product basis}, the $\kappa$- Poincar\'e commutators are
\begin{align}\label{bicross}
\begin{split}
&[P_\mu, P_\nu ]= 0, \quad [ R_i, P_0 ]= 0, \quad [ R_i,P_j ]= i\epsilon_{ijk}P_k, \\
&[ R_i, R_j ]= i\epsilon_{ijk} R_k, \quad [ R_i, N_j ]= -i\epsilon_{ijk} N_k, \\
&[N_{i},P_{0}]=iP_{i} \, \quad [N_{i},N_{j}]= -i\epsilon_{ijk}R_{k} ,\\
&[N_{i},P_{j}] = i\delta_{ij}\left(\frac{1-e^{-2\ell P_{0}}}{2\ell}+\frac{\ell}{2}{\vec{P}}^{2}\right)-i\ell P_{i}P_{j} ,
\end{split}
\end{align}
and, as a consequence, the mass Casimir gets deformed into
\begin{equation}
 \left(\frac{2}{\ell} \sinh\left(\frac{P_0}{2\ell}\right)\right)^2 - e^{\ell P_0}P_i P^i .
\end{equation}
The modification of the commutators between boosts and translation generators also
requires a deformation of the laws of composition of momenta. Indeed, one can easily check that under the action of $N_i$,
 give in terms of the commutators \eqref{bicross}, one would have that
\begin{equation}
[N^i_{[p,k]}, p_\mu+k_\mu] = [N^i_p + N^i_k, p_\mu + k_\mu] \neq 0 ,
\end{equation}
even if $p_\mu+ k_\mu = 0$. This means that the usual conservation laws would not be covariant.
As one can check by using the relations in Eqs. \eqref{bicross}, the correct modification (in order to achieve covariance) is
\begin{equation}
p_\mu \oplus_{\ell} k_\mu =\begin{cases} p_{0}+k_{0}\\ p_{i}+e^{- \ell p_{0}}k_{i}\end{cases} ,
\end{equation}
for momenta, while for the boosts one has
\begin{equation}
N^i_p \oplus N^i_k = N^i_p + e^{- \ell p_{0}} N^i_k + \ell\epsilon_{ijn} p^j R^n_k  .
\end{equation}

These modifications are such that now the condition of covariance of the composition law is obeyed
\begin{equation}
 [N^i_p \oplus N^i_k, p_\mu \oplus k_\mu] = 0 ,
\end{equation}
as the reader can easily verify.

In the formalism of Hopf algebras these observations can be formalized by saying that the co-algebra, i.e. the set of relations that define the action of the generators on the product of fields, is non-primitive and, in particular, is modified as follows
\begin{equation}
\Delta P_i = P_i \otimes 1 + e^{-\ell P_0} \otimes P_i \, , \quad \Delta N_i = N_i \otimes 1 + e^{-\ell P_0} \otimes N_i + \ell \epsilon_{ijk} P_j \otimes R_k
\end{equation}
 The coproducts of $R^i$ and $P_0$ have not been written down explicitly because they remain primitive, i.e. they are dictated by the Leibniz rule. \\
To sum up, we have seen that there are two key ingredients needed by a deformed relativistic picture, i.e. the deformed algebra closed by the generators of non-linearly deformed symmetry transformations has to be compatible with both the form of the mass Casimir (or, equally, the associated on-shell relation) and the conservation laws for the associated charges. We stress that for compatibility we mean that the boost generator must leave invariant the on-shell relation and must transform covariantly the composition law for momenta.

\section{Mixing coproduct: $\kappa$-Poincar\'{e} case of study}
\label{sec:mixingl-l'}

We are now ready to contemplate a further generalization of the notion of relativistic-symmetry deformation by allowing for \textit{non-universal} scenarios.
In this Section we introduce the notion of \textit{mixing coproduct} and provide a suitable formalization of it. As already mentioned, this would allow us to formulate a relativistic model where different particles (i.e. particles with different quantum numbers) do not follow the same relativistic laws. In particular, within the formalism of Hopf algebras, we can infer how to compose particles' four momenta from the coalgebraic sector. Thus, the mixing coproduct would be a mathematical object that gives us ``mixed" composition laws where the momenta we compose represent the charges associated to translation generators belonging to different algebras (for instance a Lie algebra and a Hopf algebra, or two different Hopf algebras) or, in some cases, to different ``bases" of the same Hopf algebra. We shall explain in this section what we mean exactly by these different cases.

Let us start by considering three Hopf algebras $H$, $H'$ and $H''$ and two maps $\phi$ and $\phi'$ defined as
\begin{equation}
\phi:H \to H'', \qquad \phi':H'\to H''
\end{equation}
 with inverse given by $\phi^{-1}$ and $\phi'^{-1}$ respectively. Then, it is possible to define a mixing coproduct \cite{gacmixingold} by composing these two maps as follows
\begin{align}\label{mixing coproduct}\begin{split}
&H'' \xrightarrow[]{\Delta''} H''\otimes H'' \xrightarrow[]{\phi^{-1} \otimes \phi'^{-1}} H\otimes H'.\\
\end{split}\end{align}

Thanks to the mixing coproduct $\phi^{-1} \otimes \phi'^{-1}$, we are here composing the momenta of two particles, whose symmetries are dictated by $H$ and $H'$ respectively, and the resulting particle with momentum given by the sum follows again a distinct symmetry group, i.e. $H''$. (i.e. it is the corresponding charge of the generator of translations in $H''$). It is not difficult to realize that, introducing analogous maps, it would be possible to have also mixing coproducts with target space either $H \otimes H''$ or $H' \otimes H''$. \\
In order to gather some confidence with this novel object and the related formalism, we shall discuss a couple of relevant examples. \\
As first example we consider the case in which $H,H',H''$ are three $\kappa$-Poincar\'{e} algebras $\kappa\mathcal{P}$, $\kappa'\mathcal{P}$ and $\kappa''\mathcal{P}$ which differ for the magnitude of the deformation parameter, $\ell$, $\ell'$, and $\ell''$ respectively. By means of the maps $\phi$ and $\phi'$, which in this case are simply morphisms, we can define two different ways to compose momenta in a mixed way
$$\oplus_{l\ell'}:M\oplus M' \to M'',$$
$$\oplus_{\ell'l}:M'\oplus M \to M'',$$
where here $M$, $M'$ and $M''$ stand for the three momentum spaces. This can be done by writing down explicitly the actions of the morphisms $\phi$ and $\phi'$ over the algebras $\kappa\mathcal{P}$ and $\kappa'\mathcal{P}$ respectively. \\
It is known that a given Hopf algebra can be written, as far as explicit formulas are concerned, in some rather different ways, depending on the conventions adopted \cite{kowalski2003non}. In fact, for Hopf algebras one must allow both linear and non-linear maps between the generators, mapping one ``basis" into another \cite{kowalski2003non}.

If we express the three Hopf Algebras all in the bicrossproduct basis, then the simplest way to define the morphisms $\phi$ and $\phi'$ is given by the following expressions
\begin{equation}\label{isol-l'}
\begin{alignedat}{3}
&\phi(P_{\mu})={\ell'' \over \ell }P_{\mu}'', \qquad &&\phi'(P_{\mu}')={\ell'' \over \ell'}P_{\mu}'', \\
&\phi(R_{i})=R_{i}'', \qquad &&\phi'(R_{i}')=R_{i}'', \\
&\phi(N_{i})=N_{i}'', \qquad &&\phi'(N_{i}')=N_{i}''.
\end{alignedat}
\end{equation}
Here $G \, \in \, \kappa\mathcal{P}$, $G' \, \in \, \kappa'\mathcal{P}$, and $ G'' \, \in \, \kappa''\mathcal{P}$ are used to denote the symmetry generators (in the bicrossproduct basis) of the three $\kappa$-Poincar\'{e} algebras characterized by different deformation parameters $\ell$, $\ell'$, and $\ell''$. In particular, it is possible to prove that these maps define isomorphisms between the Hopf algebras. For the sake of brevity, let us focus only on the morphism $\phi$. The reader can straightforwardly verify that
\begin{align*}
&\phi([P_{\mu},P_{\nu}])=0=[\phi(P_{\mu}),\phi(P_{\nu})],\\
&\phi([R_i,R_j])=\epsilon_{ijk}R_{k}''=[\phi(R_i),\phi(R_j)],\\
&\phi([N_i,N_j])=-\epsilon_{ijk}R_{k}''=[\phi(N_i),\phi(N_j)],\\
&\phi([R_i,N_j])=\epsilon_{ijk}N_k''=[\phi(R_i),\phi(N_j)] .\\
&\phi([R_i,P_0])=0=[\phi(R_i),\phi(P_0)],\end{align*}
It is also rather simple to verify the following equalities
\begin{align*}
&\phi([R_i,P_j])=\epsilon_{ijk}{\ell'' \over \ell }P_k''={\ell'' \over \ell }[R_i'',P_j'']=[\phi(R_i),\phi(P_j)],\\
&\phi([N_i,P_0])={\ell'' \over \ell }P_i''={\ell'' \over \ell }[N_i'',P_0'']=[\phi(N_i),\phi(P_0)]
\end{align*}
and finally
 \begin{align*}
 \phi([N_i,P_j]) &=\phi\Bigl[\delta_{ij}\Bigl({1-e^{-2 \ell P_{0}} \over 2\ell}+{\ell \over 2}|\bar{P}|^2\Bigr)- \ell P_{j}P_{k}\Bigr]\\
 &=\delta_{ij}\Bigl({1-e^{ -2 \ell { \ell'' \over \ell }P_{0}''} \over 2 \ell}+{\ell \over 2}{\ell''^2 \over \ell^2}|\bar{P}''|^2\Bigr)- \ell {\ell''^2 \over \ell^2}P_{j}''P_{k}''\\&={\ell'' \over \ell}
 \Bigl[\delta_{ij}\Bigl({1-e^{-2\ell''P_{0}''} \over 2\ell''}+{\ell'' \over 2} \ell \bar{P}''|^2\Bigr)-\ell''P_{j}''P_{k}''\Bigr]\\&=[\phi(N_i),\phi(P_j)].
 \end{align*}
These observations establish the isomorphism at the level of the algebra sector (i.e. commutators). Then, we need to look also at the coalgebra. For the coproducts we indeed find
\begin{align*}
 & \phi \otimes \phi ( \Delta ( P_0 ) ) = { \ell'' \over \ell } P_0 '' \otimes \mathds{1} + \mathds{1} \otimes { \ell'' \over \ell} P_0 = \Delta'' ( \phi ( P_0 )) , \\
 &\phi\otimes\phi(\Delta(P_i))={\ell'' \over \ell} P_0'' \otimes\mathds{1}+e^{-\ell{\ell'' \over\ell }P_0}\otimes{\ell'' \over \ell}P_0=\Delta''(\phi(P_i)),\\
 &\phi\otimes\phi(\Delta(R_i))=R_i''\otimes\mathds{1}+\mathds{1}\otimes R_i''=\Delta''(\phi(R_i))
\end{align*}
and
\begin{align*}
\phi\otimes\phi(\Delta(N_i)) & =N_i''\otimes\mathds{1}+e^{-\ell{\ell'' \over \ell}P_0}\otimes N_i''+\ell\epsilon_{ijk}{\ell'' \over \ell}P_j''\otimes R_k'' = \Delta''(\phi(N_i)).
\end{align*}
The last check we need concerns the compatibility of the map with the antipodes (then the compatibility with the counits follows straightforwardly), and also in this case it is easy to verify that
\begin{align*}
&\phi(S(P_0))=-{\ell'' \over \ell}P_0''=S''(\phi(P_0)),\\
&\phi(S(P_i))=-e^{-\ell{\ell'' \over \ell}P_0''}{\ell'' \over \ell}P_i''=S''(\phi(P_i)),\\
&\phi(S(R_i))=-R_i''=S''(\phi(R_i))\\
\end{align*}
and finally
$$\phi(S(N_i))=-e^{ \ell {\ell'' \over \ell }P_0''}N_i''+ \ell \epsilon_{ijk}e^{ \ell {\ell'' \over\ell }P_0''}{\ell'' \over\ell }P_j''R_k''=S''(\phi(N_i)).$$

We therefore established that the morphisms $\phi$ and $\phi'$ are actually isomorphisms connecting $\kappa\mathcal{P}$ with $\kappa' \mathcal{P}$ and $\kappa'\mathcal{P}$ with $\kappa'' \mathcal{P}$. These isomorphisms also has an inverse map, which for example
for $\phi$ is given by
\begin{align*}
&\phi^{-1}(P_{\mu}'')={ \ell \over \ell''}P_{\mu}, \\
&\phi^{-1}(R_{i}'')=R_{i}, \\
& \phi^{-1}(N_{i}'')=N_{i} .
\end{align*}
Of course, the inverse also constitutes an isomorphism between Hopf algebras. Indeed inverting the isomorphism
 simply amounts to exchange of roles between deformation scales, for example exchanging the roles of $\ell$ and $\ell''$. \\
Using the two morphisms $\phi$ and $\phi'$ we can construct, as anticipated in \eqref{mixing coproduct},
the mixing coproducts involving these three $\kappa$-Poincar\'{e} algebras as follows
\begin{align}\label{isol-l'2}\begin{split}
&p \oplus_{\ell\ell'}'' q=\begin{cases}{\ell'' \over\ell }p_{0}+{\ell'' \over \ell'}q_{0}\\{\ell'' \over\ell }p_{i}+{\ell'' \over \ell'}e^{- \ell p_{0}}q_{i}\end{cases},\\
&q \oplus_{\ell'\ell}'' p=\begin{cases}{\ell'' \over \ell'}q_{0}+{\ell'' \over\ell }p_{0}\\ {\ell'' \over \ell'}q_{i}+{\ell'' \over\ell }e^{-\ell'q_{0}}p_{i}\end{cases} .\\
\end{split}\end{align}

These relations give us two possible composition laws between momenta that belong to different momentum spaces. In fact, here $p\in M$, $q\in M'$ while, by definition,
$p \oplus_{l\ell'}'' q , \, q \oplus_{\ell' \ell}'' p \, \in M''$.
Consequently, this simple framework provides us a first example of a deformed relativistic theory where we are able to compose particles' momenta that live on different momentum spaces or, in other words, represent the charges associated to the symmetry transformations of different Hopf algebras. These two sums of momenta \eqref{isol-l'2} differ only for the order of the addenda.

 It should be noticed that these laws are not well defined when one of the three deformation parameters vanishes, i.e. we have to impose that $\ell \neq 0 \,$ , $\ell' \neq 0 \,$ , $\ell'' \neq 0$. This means that the above introduced morphisms cannot be used to compose a particle with symmetries described by the $\kappa$-Poincar\'{e} group with another whose momenta follow the standard Poincar\'{e} symmetries. As we shall see later, within our framework, this can be done with another class of maps. However, let us point out that this
 is consistent with the fact that we have proven they are isomorphisms between Hopf algebras and, therefore, they could not relate the Poincaré Lie algebra with a Hopf algebra since there is no isomorphism connecting them. Finally, let us notice that, according to \eqref{isol-l'2}, given two momenta $p$ and $q$ of two particles in $\kappa\mathcal{P}$ and in $\kappa'\mathcal{P}$ respectively, then the system of equations represented by the condition $p\oplus q=0$ (i.e. the condition of the conservation of momenta) does not depend on the specific choice of the target momentum space $M''$ nor on the order of addenda.

A second interesting example that allows us to study the properties as well as the meaning of mixing coproducts has been first studied in Ref.\cite{barcaroli2014relative}. In this case, one considers the map
\begin{equation}
\psi:\kappa{\mathcal{P}}\to\kappa{\mathcal{P}} ,
\end{equation}
which is explicitly given by the following formulas
\begin{align}\begin{split}\label{mappaLeo}
&P_0=\psi(K_0)=K_0,\\
&P_i=\psi(K_i)=e^{{ \ell \over 2}K_0}K_i,\\
&R_{i}=\psi(M_{i})=M_i,\\
&N_{i}=\psi(B_i)=e^{{ \ell \over 2}K_0}(B_i-{ \ell \over 2}\epsilon_{ijk}K_jM_k),
\end{split}\end{align}
where $(P_\mu , R_i, N_j)$ are the generators of the $\kappa$-Poincaré algebra in the bicrossproduct basis while $(K_\mu, M_i, B_i)$ are the generators of the $\kappa$-Poincaré algebra in the so-called classical basis \cite{kosinski1994classical}. In the classical basis we have
\begin{align}\label{class}
\begin{split}
&[B_i, K_0] = iK_i \, , \quad [B_i, K_j] = i\delta_{ij}K_0 \, , \quad [B_i, B_j] = -i \epsilon_{ijk}M_k , \\
&[M_i, K_0] = 0 \, , \quad [M_i, K_j] = i\epsilon_{ijk}K_k \, , \quad [M_i, M_j] = i \epsilon_{ijk}M_k , \\
&[K_\mu, K_\nu] = 0 \, , \quad [M_i, B_j] = i\epsilon_{ijk}B_k ,
\end{split}
\end{align}
while the coproducts are rather complicated and lengthy, so we do not report them but they can be found in \cite{Borowiec} or in references therein. \\
Thanks to this map $\psi$ and following steps similar to those we did above, one can write for instance the composition law
\begin{align}\begin{split}\label{mixingLeo}
&p \oplus_{B-S}^B k=\begin{cases}p_{0}+k_{0}\\p_{i}+e^{- \ell p_{0}}e^{{\ell \over 2}k_0}k_{i}\end{cases}\\
\end{split}\end{align}
that mixes particles $p$ and $k$ obeying the same symmetry group but expressed in two different bases (bicrossproduct the former and standard the latter), while giving back a momentum in the $\kappa$-Poincar\'{e} bicrossproduct basis.

While in the first part of this section we "mixed" Hopf algebras with different deformation parameter but described in the same basis,
here we are "mixing" two copies
of the same Hopf algebra (same deformation parameter) but adopting two different bases.
From a phenomenological point of view, one of the main points of interest resides in the fact that in the bicrossproduct basis one has on-shellness of the type
\begin{eqnarray}
{4 \over \ell^2}\sinh^2\Bigl({\ell \over 2}p_0\Bigr) = e^{\ell p_0}p_i p^i
\end{eqnarray}
whereas for the classical basis the on-shellness is undeformed
\begin{eqnarray}
k^2_0 = k_i k^i .
\end{eqnarray}

\section{Mixing coproduct for the Magueijo-Smolin DSR picture}
\label{MS}

Before moving on to more ambitious implementations of the notion of mixing coproduct,
we find appropriate to offer a brief aside showing that the mixing-coproduct techniques introduced
in the previous section
can also shed light on results previously obtained without abstracting the notion of mixing coproduct.
Our main objective for this small aside it to show that a composition law introduced
by Magueijo and Smolin in Ref.\cite{magueijo2003generalized} can be rephrased in the language of mixing coproducts. The key
ingredient is an operator
$$U[\ell]\equiv \exp(\ell P_0 D), $$
acting on the Poincar\'{e} algebra, with
$$D\equiv P_0{\partial \over \partial P_0}+P_1{\partial \over \partial P_1}.$$
Then, taking the generators transformed by this map
$$X[\ell]\equiv U^{-1} X U,$$
one can easily check that $X[\ell]$ still obey the Poincar\'{e} commutation rules. Thus, we have again the Poincar\'{e} algebra but with a modified coalgebric sector and with generators of infinitesimal translations given by
$$P_i[\ell]\equiv U[\ell] P_i = {P_i \over 1+\ell P_0},$$
and its inverse
$$P_i={P_i[\ell] \over 1-\ell P_0[\ell]}.$$

We also define the map $u[\ell]$
$$X \mapsto U X[\ell] U^{-1},$$
which for the generators $P_i$ reads
$$P_i \mapsto {P_i[\ell] \over 1-\ell P_0[\ell]}.$$

Following the formalism used in the previous section for the mixing coproduct, we construct the chain of maps
\begin{equation}
\label{cat1}
P\Bigl[{\ell \over 2}\Bigr]\xrightarrow{u^{-1}\bigl[{\ell \over 2}\bigr]}P\xrightarrow{\Delta}P\otimes P\xrightarrow{u[\ell]\otimes u[\ell]}P[\ell]\otimes P[\ell],
\end{equation}
which, when applied backwards, gives the momentum space the Magueijo-Smolin composition relation \cite{magueijo2003generalized}
$${p_i \over 1- {\ell \over 2} p_0}={q_i \over 1-\ell q_0}+{k_i \over 1-\ell k_0},$$
i.e.,
\begin{equation}
p_i={q_i+k_i-\ell q_0 k_i-\ell q_i k_0 \over 1-{\ell \over 2}(q_0+k_0)}.
\label{eq:MS_composition_rule}
\end{equation}

As a further aside it is amusing to notice that there is also another route for obtaining the Maueijo-Smolin
composition law. We start doing that by noticing that we already have a map linking $P\bigl[{\ell \over 2}\bigr]$ to $P[\ell]$:
$$P\Bigl[{\ell \over 2}\Bigr]\xrightarrow{u^{-1}\bigl[{\ell \over 2}\bigr]}P\xrightarrow{u[\ell]}P[\ell].$$
We call this map $w\bigl[{\ell \over 2},\ell\bigr]$
\begin{equation}
\label{w}
w\Bigl[{\ell \over 2},\ell\Bigr] \triangleright P_{i}\Bigl[{\ell \over 2}\Bigr] = {P_{i}[\ell] \over 1-{\ell \over 2}P_0[\ell]}.
\end{equation}

Now we need to prove that the chain in Eq. \eqref{cat1} is indeed the same chain given by
\begin{equation}
\label{cat2}
P\Bigl[{\ell \over 2}\Bigr]\xrightarrow{\Delta\bigl[{\ell \over 2}\bigr]}P\Bigl[{\ell \over 2}\Bigr]\otimes P\Bigl[{\ell \over 2}\Bigr]\xrightarrow{w\bigl[{\ell \over 2},\ell\bigr]\otimes w\bigl[{\ell \over 2},\ell\bigr]}P[\ell]\otimes P[\ell].
\end{equation}

First we observe that the coproduct of $P_i\bigl[{\ell \over 2}\bigr]$ can be obtained using the relations
$$P_i\Bigl[{\ell \over 2}\Bigr]={P_i \over 1+{\ell \over 2}P_0}, \quad P_i={P_i[{\ell \over 2}] \over 1-{\ell \over 2}P_0[{\ell \over 2}]},$$
from which it follows that
\begin{align}
\label{cop}
\begin{split}
\Delta\Bigl[{\ell \over 2}\Bigr]P_i\Bigl[{\ell \over 2}\Bigr]= & {P_i \otimes \mathbb{I}+\mathbb{I}\otimes P_i \over 1+{\ell \over 2}(P_0 \otimes \mathbb{I}+\mathbb{I}\otimes P_0)} \\
= & {{P_i\bigl[{\ell \over 2}\bigr] \over 1-{\ell \over 2}P_0\bigl[{\ell \over 2}\bigr]} \otimes \mathbb{I}+\mathbb{I}\otimes {P_i\bigl[{\ell \over 2}\bigr] \over 1-{\ell \over 2}P_0\bigl[{\ell \over 2}\bigr]} \over 1+{\ell \over 2}({P_0\bigl[{\ell \over 2}\bigr] \over 1-{\ell \over 2}P_0\bigl[{\ell \over 2}\bigr]} \otimes \mathbb{I}+\mathbb{I}\otimes {P_0\bigl[{\ell \over 2}\bigr] \over 1-{\ell \over 2}P_0\bigl[{\ell \over 2}\bigr]})}.
\end{split}
\end{align}

At this point given two particles with momenta $q$ and $k$ respectively in $P[\ell]$ and applying backwards the chain in Eq. \eqref{cat2}, we obtain the transformation
$$(q_i,k_i) \mapsto \Bigl({q_i \over 1-{\ell \over 2}q_0},{k_i \over 1-{\ell \over 2}k_0}\Bigr)$$
which has to be replaced in the coproduct of Eq. \eqref{cop}.
In order to do so we first notice that if
$$x_i\mapsto {x_i \over 1-{\ell \over 2}x_0}$$
then
$${x_i \over 1-{\ell \over 2} x_0} \mapsto {{x_i \over 1-{\ell \over 2}x_0} \over 1-{\ell \over 2}{x_0 \over 1-{\ell \over 2}x_0}}={x_i \over 1-\ell x_0},$$
Thus, substituting $\Bigl({q_i \over 1-{\ell \over 2}q_0},{k_i \over 1-{\ell \over 2}k_0}\Bigr)$ in the coproduct of Eq. \eqref{cop} we have
\begin{align*}
{{q_i \over 1-\ell q_0}+{k_i \over 1-\ell k_0} \over 1+{\ell \over 2}\Bigl({q_0 \over 1-\ell q_0}+{k_0 \over 1-\ell k_0}\Bigr)} ={q_i+k_i-\ell q_0 k_i-\ell q_i k_0 \over 1-{\ell \over 2}(q_0+k_0)},
\end{align*}
which is exactly the Magueijo-Smolin composition rule of Eq. \eqref{eq:MS_composition_rule}.

The map $w\bigl[{\ell \over 2},{\ell}\bigr]:P\bigl[{\ell \over 2}\bigr]\to P[\ell]$
is indeed like a rescaling map between two $\kappa$-Poincar\'{e} algebra with two distinct deformation scales $\ell$ and ${\ell \over 2}$, i.e.,
$${P_i\bigl[{\ell \over 2}\bigr] \over 1-{\ell \over 2}P_0\bigl[{\ell \over 2}\bigr]}\mapsto{{P_{i}[\ell] \over 1-{\ell \over 2}P_0[\ell]} \over 1-{\ell \over 2}{P_{0}[\ell] \over 1-{\ell \over 2}P_0[\ell]}}={P_{i}[\ell] \over 1-\ell P_0[\ell]}.$$

The equivalance between the chains in Eq. \eqref{cat1} and \eqref{cat2} can also be interpreted as the equivalence between maps
$$\Delta\circ u^{-1}[\ell]=(u^{-1}[\ell]\otimes u^{-1}[\ell])\circ \Delta[\ell],$$
which shows the compatibility between the coproducts of $P$ and $P[\ell]$ algebras, and more generally between the coproducts of $P[\ell']$ and $P[\ell]$ algebras.

\section{Mixing coproducts between Poincar\'{e} and $\kappa$-Poincar\'{e} algebras}
\label{sec:classificazione}

So far we focused on coproducts mixing pairs of algebras that were isomorphic to one another (or two bases of the same algebra). In this
section we show that one can consistently introduce mixing coproducts also for non-isomorphic algebras,
focusing on the case of mixing the standard Poincar\'{e} (Lie) algebra and the $\kappa$-Poincar\'{e} Hopf algebra.
 It should be noticed that this is rather challenging even though the Poincar\'{e} (Lie) algebra
 is obtained from the $\kappa$-Poincar\'{e} Hopf algebra in the limit in which the deformation parameter is removed.
 In fact, the mixing coproducts we analyzed in Section III all involve maps which are not analytic as the deformation parameters are removed ($\ell \, \rightarrow 0$). In this section we shall truly need a new type of mixing coproduct. For definiteness and simplicity we
 focus on the case of a 1+1-dimensional spacetime. 

 We start by introducing notation for the most general composition law\footnote{Here and in the rest of this paper, if not otherwise specified, we denote with $\boxplus$ the mixing coproducts obtained following this procedure.}
\begin{equation}
p \boxplus k = \begin{cases} \epsilon(p_0,p_1,k_0,k_1)p_0+\zeta(p_0,p_1,k_0,k_1)k_0 \\ f(p_0,p_1,k_0,k_1)p_1+g(p_0,p_1,k_0,k_1)k_1 \end{cases},
\label{eq:comp1}
\end{equation}
where $p$ is the momentum of a $\kappa$-Poincar\'{e} particle and $k$ is the momentum of a Poincar\'{e} particle\footnote{By stating
that $p$ is the momentum of a $\kappa$-Poincar\'{e} particle we mean that if $p'$ is another particle momentum of the same type, then
$$p\oplus p'=\begin{cases}p_0+p_0',\\p_1+e^{-\ell p_0}p_1'\end{cases},$$
whereas of course if $k$ and $k'$ are two Poincar\'{e} particle momenta then obviously
$$k+k'=\begin{cases}k_0+k_0',\\k_1+k_1'\end{cases}.$$}, and we require that under a suitable boost generator $N_{[p,k]}$
\begin{equation}\label{conservazione_quadrimpulso}p \boxplus k=0 \Rightarrow [N_{[p,k]},p \boxplus k]=0.\end{equation}

This requirement guarantees the covariance of the composition rule in \eqref{eq:comp1}. In \eqref{eq:comp1} $\epsilon$, $\zeta$, $f$ and $g$ are general functions of the momenta $p,k$, which for $\ell \to 0$ have to be equal to 1.
We shall consistently denote by $p$ momenta of the $\kappa$-Poincar\'{e} type and by $k$ momenta of the Poincar\'{e} type.

Firstly, let us focus on two ``complementary" proposals which are respectively
\begin{equation}\label{sistemaEp0,p1}p \boxplus k=\begin{cases}\varepsilon(p_0,p_1)p_0+k_0 \\ f(p_0,p_1)p_1+k_1\end{cases}\end{equation}
and
\begin{equation}\label{sistemaEk0,k1}p \boxplus k=\begin{cases}p_0+\tilde{\varepsilon}(k_0,k_1)k_0 \\ p_1+\tilde{f}(k_0,k_1)k_1\end{cases},\end{equation}
where the boost must be of the form $N_{[p,k]}=h(p_0,p_1)N_{[p]}+N_{[k]}$ for the former case, and $N_{[p,k]}=N_{[p]}+\tilde{h}(k_0,k_1)N_{[k]}$ for the latter.

As we explicitly show in Appendix B, given the compatibility condition \eqref{conservazione_quadrimpulso} all the solutions of these systems, i.e. \eqref{sistemaEp0,p1} and \eqref{sistemaEk0,k1}, are of the type
\begin{equation}\label{box eps p0p1}p \boxplus k=\begin{cases} \varepsilon p_0+k_0 \\\sqrt{\varepsilon^2p_0^2-{4 \over \ell^2}\sinh^2\Bigl({\ell\over 2}p_0\Bigr)+e^{ \ell p_0}p_1^2}{p_1 \over | p_1|}+k_1\end{cases}\end{equation}
and
\begin{equation}\label{box eps k0k1}p \boxplus k=\begin{cases} p_0+\tilde{\varepsilon}k_0 \\\ p_1+e^{{\ell \over 2}\tilde{\varepsilon}k_0}\sqrt{{4 \over \ell^2}\sinh^2\Bigl({\ell \over 2}\tilde{\varepsilon}k_0\Bigr)-k_0^2+k_1^2}{k_1 \over | k_1|}\end{cases}\end{equation}
respectively. Thus, we would be left with just one undetermined function: $\varepsilon$ for the first system, and $\tilde{\varepsilon}$ for the second.

However, it is rather easy to understand that not all the choices for $\varepsilon$ and $\tilde{\varepsilon}$ can be acceptable. Indeed, if we take a look at Eqs. \eqref{box eps p0p1} and \eqref{box eps k0k1}, then it follows that we must have \footnote{Notice that these two disequalities can be obtained by imposing zero spatial momentum either for the particle $p$ or for the particle $k$ respectively.}
\begin{equation}
\varepsilon p_0^2 \ge {4 \over \ell^2}\sinh^2\Bigl({\ell \over 2}p_0\Bigr) \quad \to \quad \varepsilon \ge {2 \over \ell p_0}\sinh\Bigl({ \ell \over 2}p_0\Bigr) ,
\end{equation}
for the first case, and
\begin{equation}
{4 \over \ell^2}\sinh^2\Bigl({\ell \over 2}\tilde{\varepsilon}k_0\Bigr)\ge k_0^2 \quad \to \quad \tilde{\varepsilon} \ge {2 \over \ell}\ln\Bigl({\ell \over 2}k_0+\sqrt{{\ell^2 \over 4}k_0^2+1}\Bigr)
\end{equation}
in the second situation. Essentially, this is a direct consequence of the functional form of the expressions. Moreover, it is natural to require that $\varepsilon$ and $\tilde{\varepsilon}$ have to be both surjective and injective for any value of, respectively, $p_1$ and $k_1$. They must also go to $1$ for $\ell$ going to zero and, finally, need to have the same sign of the energy ($p_0$ or $k_0$) in order to guarantee that the above introduced mixing coproducts actually represent a good and consistent choice for composing the momenta (the energies) of two particles.

Given that, we can now deduce several facts about the mixing coproducts we are trying to construct. For instance, if we concentrate on Eq. \eqref{box eps p0p1} and consider the case in which $\varepsilon$ depends only on $p_0$ and enjoys the aforementioned properties, then $\varphi(p_0)=\varepsilon p_0$ will have an inverse $\tilde{\varphi}(k_0)$ with the same characteristics. We can then choose
 $\tilde{\varepsilon}={\tilde{\varphi} \over k_0}$ in Eq. \eqref{box eps k0k1}. Notice that, with these choices, when we impose momentum conservation in a scattering process (i.e. $p\boxplus (-k)=0$), both composition laws give the same relations between $p$ and $k$. From the former we find
\begin{equation}
p \boxplus (-k)=0 \quad \to \quad \begin{cases}k_0=\varphi(p_0) \\ k_1=\sqrt{\varphi^2(p_0)-{4 \over \ell^2}\sinh^2\Bigl({\ell \over 2}p_0\Bigr)+e^{\ell p_0}p_1^2}{p_1 \over | p_1|}\end{cases}
\end{equation}
and
\begin{equation}
p\boxplus (-k)=0 \quad \to \quad \begin{cases}\tilde{\varphi}(k_0)=p_0\\ k_1=\sqrt{k_0^2-{4 \over \ell^2}\sinh^2\Bigl({\ell \over 2}\tilde{\varphi}(k_0)\Bigr)+p_1^2e^{\ell \tilde{\varphi}(k_0)}}{p_1 \over |p_1|}\end{cases}
\end{equation}
from the latter. The fact that we obtained two identical systems suggests that, as we shall show later, these two composition laws can be regarded one as the inverse of the other. This is a remarkable feature since for the consistency of this composition laws we must always have
\begin{align*}&\varphi\equiv\varepsilon p_0 \ge {2 \over \ell}\sinh\Bigl({\ell \over 2}p_0\Bigr),\\
& \tilde{\varphi}\equiv\tilde{\varepsilon} k_0 \ge {2 \over \ell}\ln\Bigl({\ell \over 2}k_0+\sqrt{{\ell^2 \over 4}k_0^2+1}\Bigr).\end{align*}

From the former of the above disequalities we find
\begin{equation}
\varphi^{-1}\varphi(p_0)=\tilde{\varphi}\varphi(p_0)=p_0 \ge \tilde{\varphi}\Bigl[{2 \over \ell}\sinh\Bigl({\ell \over 2}p_0\Bigr)\Bigr],
\end{equation}
or equivalently
\begin{equation}
\tilde{\varphi}(k_0) \le {2 \over \ell}\ln\Bigl({\ell \over 2}k_0+\sqrt{{\ell^2 \over 4}k_0^2+1}\Bigr),
\end{equation}
where we have introduced the notation $k_0={2 \over \ell}\sinh\Bigl({\ell \over 2}p_0\Bigr)$. Due to the properties we imposed on the functions, the same relation must hold also when we exchange the two sides and thus
\begin{equation}
\tilde{\varphi}(k_0) = {2 \over \ell}\ln\Bigl({\ell \over 2}k_0+\sqrt{{\ell^2 \over 4}k_0^2+1}\Bigr)
\end{equation}
and also
\begin{equation}
\varphi(p_0)={2 \over \ell}\sinh\Bigl({\ell \over 2}p_0\Bigr).
\end{equation}

With this procedure we thus conclude that there is only \textit{one} possible pair of mixing coproducts that deforms the momenta and verifies a sort of inverse relation. This is given by
\begin{equation}\label{(p+k)P p0}p\boxplus_{\mathcal{P}}k=\begin{cases}{2 \over \ell}\sinh\Bigl({\ell \over 2}p_0\Bigr)+k_0 \\ e^{{\ell \over 2}p_0}p_1+k_1\end{cases}\end{equation}
and
\begin{equation}\label{bla}p\boxplus k=\begin{cases}p_0+{2 \over \ell}\ln\Bigl({\ell \over 2}k_0+\sqrt{{\ell^2 \over 4}k_0^2+1}\Bigr) \\ p_1+\Bigl({\ell \over 2}k_0+\sqrt{{\ell^2 \over 4}k_0^2+1}\Bigr)k_1\end{cases}.\end{equation}

Let us now consider a particle of mass $m$ with momentum $p$ whose symmetries are described by the $\kappa$-Poincaré algebra. If we define
\begin{equation}
E={2 \over \ell}\sinh\Bigl({\ell \over 2}p_0\Bigr), \qquad \Pi=e^{{\ell \over 2}p_0}p_1
\end{equation}
then
\begin{equation}
E^2-\Pi^2={4 \over \ell^2}\sinh^2\Bigl({\ell \over 2}p_0\Bigr)-e^{\ell p_0}p_1^2=m^2 .
\end{equation}

This tells us that the map $(p_0,p_1)\to(E,\Pi)$ ``transforms'' the particle $p$ into a particle with standard Poincaré symmetries\footnote{Notice that this is true also for the mixing coproducts of Eq. \eqref{box eps p0p1}.}. Given that, it is possible to regard the mixing coproduct of Eq. \eqref{(p+k)P p0} as a trivial sum of momenta once we deform the momentum coordinates $(p_0 , p_1)$, just as we saw already in the previous sections. On the other hand, we can not give a similar interpretation for the composition law in Eq.\eqref{bla}. In other words, we would like to have a sort of complementary map that transforms the momentum of a particle with Poincaré symmetries into the momentum associated to the $\kappa$-Poincaré algebra. This is not possible since if $k$ is a Poincaré particle
with mass $m$ and we define
\begin{equation}
\tilde{E}={2 \over \ell}\ln\Bigl({\ell \over 2}k_0+\sqrt{{\ell^2 \over 4}k_0^2+1}\Bigr), \qquad \tilde{\Pi}=\Bigl({\ell \over 2}k_0+\sqrt{{\ell^2 \over 4}k_0^2+1}\Bigr)k_1
\end{equation}
then
$${4 \over \ell^2}\sinh^2\Bigl({\ell \over 2}\tilde{E}\Bigr)-e^{\ell \tilde{E}}\tilde{\Pi}^2 \ne m^2,$$
and the spatial part of the coproduct \eqref{bla} is not of the form $p_1+e^{-\ell p_0}\tilde{\Pi}$, regardless of the choice of $\tilde{\Pi}$. However, we can show that a suitable modification of Eq. \eqref{bla} actually allows to overcome these obstructions and to have a composition law that admits an interpretation analogous to
Eq. \eqref{(p+k)P p0}. To this end, let us consider a general mixing coproduct of the form
\begin{equation}\label{p0k0}p\boxplus k=\begin{cases}\varepsilon(p_0,k_0)p_0+k_0 \\ f(p_0,k_0)p_1+k_1\end{cases},\end{equation}
with $N_{[p,k]}=hN_{[p]}+N_{[k]}.$ Then we have
\begin{equation}
[N_{[p,k]},\varepsilon p_0+k_0]=h\Bigl({\partial \varepsilon \over \partial p_0}p_1p_0+\varepsilon p_1\Bigr)+{\partial \varepsilon \over \partial k_0}k_1p_0+k_1
\end{equation}
and
\begin{equation}
[N_{[p,k]},fp_1+k_1]=h\Bigl[{\partial f \over \partial p_0}p_1^2+f\Bigl({1-e^{-2 \ell p_0} \over 2 \ell}-{\ell \over 2}p_1^2\Bigr)\Bigr]+{\partial f \over \partial k_0}k_1p_1+k_0.
\end{equation}

Consequently $\varepsilon$ and $f$ must satisfy the following system of differential equations
\begin{equation}\label{sistema p0k0}\begin{cases}h\Bigl({\partial \varepsilon \over \partial p_0}p_1p_0+\varepsilon p_1\Bigr)-fp_1\Bigl({\partial \varepsilon \over \partial k_0}p_0+1\Bigr)=0 \\ h\Bigl[{\partial f \over \partial p_0}p_1^2+f\Bigl({1-e^{-2 \ell p_0} \over 2 \ell}-{\ell \over 2}p_1^2\Bigr)\Bigr]-f{\partial f \over \partial k_0}p_1^2-\varepsilon p_0=0 \\ \varepsilon p_0+k_0=0\end{cases}.\end{equation}

We redirect the reader to the Appendix A for the detailed analysis of this system. The solution is given by any pair of function $(f,\varepsilon)$ that reduces to Eqs. \eqref{(p+k)P p0} when $\varepsilon p_0+k_0=0$, provided that, as always, they also have the correct behavior for $\ell \rightarrow 0$ and do not present any sort of singularity. Notice that the fact that $\varepsilon={2 \over \ell p_0}\sinh\Bigl({\ell \over 2}p_0\Bigr)$
 when $\varepsilon p_0+k_0=0$ allows us to rewrite the constraints as
\begin{equation}
{2 \over \ell }\sinh\Bigl({\ell \over 2}p_0\Bigr)+k_0=0,
\end{equation}
or also
\begin{equation}
p_0+{2 \over \ell }\ln\Bigl({\ell \over 2}k_0+\sqrt{{\ell^2 \over 4}k_0^2+1}\Bigr)=0.
\end{equation}

Thus, when the constraint is satisfied, we have
\begin{equation}
\varepsilon={2 \over \ell p_0}\sinh\Bigl({\ell \over 2}p_0\Bigr)={k_0 \over {2 \over \ell }\ln\Bigl({\ell \over 2}k_0+\sqrt{{\ell^2 \over 4}k_0^2+1}\Bigr)}
\end{equation}
and
\begin{equation}
f=e^{{\ell \over 2} p_0}=e^{-\ln\Bigl({\ell \over 2}k_0+\sqrt{{\ell^2 \over 4}k_0^2+1}\Bigr)}={1 \over {\ell \over 2}k_0+\sqrt{{\ell^2 \over 4}k_0^2+1}}.
\end{equation}

Finally, defining $\tilde{\varepsilon}={1 / \epsilon}, \quad \tilde{f}={1 / f}$, we can rewrite the system of Eq. \eqref{p0k0} as
\begin{equation}\label{p0k0invmolt}p\boxplus k=\begin{cases}\varepsilon(p_0,k_0)(p_0+\tilde{\varepsilon}(p_0,k_0)k_0) \\ f(p_0,k_0)(p_1+\tilde{f}(p_0,k_0)k_1)\end{cases},\end{equation}
where now $\tilde{\varepsilon}$ and $\tilde{f}$ reduce to \eqref{bla} on the constraint surface. \\
It is rather easy to prove that, given a general mixing coproduct of the form
\begin{equation}
p\boxplus k=\begin{cases}\tilde{\zeta}(p_0+\tilde{\varepsilon}k_0) \\ \tilde{g}(p_1+\tilde{f}k_1)\end{cases}
\end{equation}
or also
\begin{equation}
p\boxplus k=\begin{cases}\zeta(\varepsilon p_0+k_0) \\ g(fp_1+k_1)\end{cases} ,
\end{equation}
then the solutions $\tilde{\varepsilon}$, $\tilde{f}$, $\varepsilon$ and $f$ do not depend on the common factors $\tilde{\zeta}$, $\tilde{g}$, $\zeta$ and $g$, respectively. In fact, considering for instance the former case and acting with a boost $N_{[p,k]}=\tilde{i}(N_{[p]}+\tilde{h}N_{[k]})$, we find that $N_{[p,k]} \triangleright (p\boxplus k)_0$ has the following form
\begin{equation}
[\tilde{i}(N_{[p]}+\tilde{h}N_{[k]})\triangleright \tilde{\zeta}](p_0+\tilde{\varepsilon}k_0)+\tilde{\zeta}\tilde{i}(N_{[p]}+\tilde{h}N_{[k]})\triangleright(p_0+\tilde{\varepsilon}k_0).
\end{equation}

If, as usual we ask that $N_{[p,k]}\triangleright p\boxplus k=0$ when $p\boxplus k=0$, then $p_0+\tilde{\varepsilon}k_0=0$ in the above equation, and as a result it is easy to realize that both $\tilde{\zeta}$ and $\tilde{i}$ do not play any role in the identification of $\tilde{\varepsilon}$ (or $\tilde{h}$). Given that, we can ignore the functions $\varepsilon(p_0,k_0)$ and $f$ in Eq. \eqref{p0k0invmolt}, which we can then rewrite as
\begin{equation}\label{p0k0inv}p\boxplus k=\begin{cases}p_0+\tilde{\varepsilon}(p_0,k_0)k_0 \\ p_1+\tilde{f}(p_0,k_0)k_1\end{cases}\end{equation}
with $\tilde{\varepsilon}={2 \over \ell k_0}\ln\Bigl({\ell \over 2}k_0+\sqrt{{\ell^2 \over 4}k_0^2+1}\Bigr)$ and $\tilde{f}={\ell \over 2}k_0+\sqrt{{\ell^2 \over 4}k_0^2+1}$, on the constraint solutions. This allows us to modify the mixing coproduct in Eq. \eqref{bla} in the way we needed. Indeed, keeping $\tilde{\varepsilon}$ unmodified with respect to the case in Eq. \eqref{bla} while changing $\tilde{f}$ as
\begin{equation}
\tilde{f}=e^{-\ell(p_0+\tilde{\varepsilon}k_0)}\Bigl({\ell \over 2}k_0+\sqrt{{\ell^2 \over 4}k_0^2+1}\Bigr)={e^{-\ell p_0} \over {\ell \over 2}k_0+\sqrt{{\ell^2 \over 4}k_0^2+1}}
\end{equation}
we eventually obtain the mixing coproduct
\begin{equation}\label{(p+k)kP_k0}p\boxplus_{\kappa\mathcal{P}} k=\begin{cases}p_0+{2 \over \ell}\ln\Bigl({\ell \over 2}k_0+\sqrt{{\ell^2 \over 4}k_0^2+1}\Bigr) \\ p_1+e^{-\ell p_0}{k_1 \over {\ell \over 2}k_0+\sqrt{{\ell^2 \over 4}k_0^2+1}}\end{cases}.\end{equation}

If we assume that $k$ is a particle with mass $m$ and define
\begin{equation}
\tilde{E}={2 \over \ell}\ln\Bigl({\ell \over 2}k_0+\sqrt{{\ell^2 \over 4}k_0^2+1}\Bigr), \qquad \tilde{f}={k_1 \over {\ell \over 2}k_0+\sqrt{{\ell^2 \over 4}k_0^2+1}}
\end{equation}
we actually find that
\begin{equation}
{4 \over \ell^2}\sinh^2\Bigl({\ell \over 2}\tilde{E}\Bigr)-e^{\ell\tilde{E}}\tilde{\Pi}^2 = m^2:
\end{equation}
i.e. the particle $k$ has been ''deformed'' into a $\kappa$-Poincaré particle. It is worth noting that Eq. \eqref{(p+k)kP_k0} can be interpreted as a non-commutative law in the sense that we can hypothesize that by exchanging $p$ with $k$ one would have to write down
\begin{equation}\label{(k+p)kP k0}k\boxplus_{\kappa\mathcal{P}} p=\begin{cases}{2 \over \ell}\ln\Bigl({\ell \over 2}k_0+\sqrt{{\ell^2 \over 4}k_0^2+1}\Bigr)+p_0 \\ {k_1 \over {\ell \over 2}k_0+\sqrt{{\ell^2 \over 4}k_0^2+1}}+{p_1 \over \Bigl({\ell \over 2}k_0+\sqrt{{\ell^2 \over 4}k_0^2+1}\Bigr)^2}\end{cases},\end{equation}
where the last term in the spatial part of the composition law can be rewritten simply as $e^{-\ell\tilde{\varepsilon}k_0}p_i$. Notice that
 we only multiplied the spatial part of the mixing coproduct \eqref{bla} by the function $e^{-\ell\tilde{\varepsilon}k_0}$, which as aforementioned does not alter the covariance of the composition law.

In summary, the composition laws in Eqs. \eqref{(p+k)P p0} , \eqref{(p+k)kP_k0} (or also \eqref{(k+p)kP k0}) define a Poincaré-like sum and a $\kappa$-Poincaré, respectively, with total momenta associated to the Poincaré or the $\kappa$-Poincaré algebra respectively.

Without the need to provide all the details of the proof (which follows the line of reasoning used to derive Eq. \eqref{p0k0}), we can generalize the above discussions to the cases
\begin{equation}
p\boxplus k=\begin{cases}\varepsilon(p_0,p_1,k_0,k_1)p_0+k_0\\f(p_0,p_1,k_0,k_1)p_1+k_1\end{cases}
\end{equation}
and
\begin{equation}
p\boxplus k=\begin{cases}p_0+\tilde{\varepsilon}(k_0,k_1,p_0,p_1)k_0\\ p_1+\tilde{f}(k_0,k_1,p_0,p_1)k_1\end{cases},
\end{equation}
where, on the constraint solutions, the above functions must coincide with either \eqref{box eps p0p1} or \eqref{box eps k0k1}.
By doing so, for any $\tilde{\varepsilon}$ we also have the non-commutative law
\begin{align*}
&p\boxplus_{\kappa\mathcal{P}}k=\begin{cases}p_0+\tilde{\varepsilon}k_0\\
p_1+e^{-\ell p_0}e^{-{\ell \over 2}\tilde{\varepsilon}k_0}\sqrt{{4 \over \ell^2}\sinh^2\Bigl({\ell \over 2}\tilde{\varepsilon}k_0\Bigr)-k_0^2+k_1^2}{k_1 \over |k_1|}\end{cases},\\
&k\boxplus_{\kappa\mathcal{P}}p=\begin{cases}\tilde{\varepsilon}k_0+p_0\\e^{-{\ell \over 2}\tilde{\varepsilon}k_0}\sqrt{{4 \over \ell^2}\sinh^2\Bigl({\ell \over 2}\tilde{\varepsilon}k_0\Bigr)-k_0^2+k_1^2}{k_1 \over |k_1|}+e^{-\ell \tilde{\varepsilon}k_0}p_1\end{cases}.\end{align*}
This concludes the classification of all the mixing coproducts involving the Poincaré and the $\kappa$-Poincaré algebras, which are of the form
\begin{equation}
p\boxplus k=\begin{cases}\varepsilon p_0+\tilde{\varepsilon} k_0\\ f p_1+\tilde{f}k_1\end{cases}.
\end{equation}

\section{Summary and outlook}

The main goal of this work was to provide illustrative examples of 
consistent scenarios with non-universal deformation of relativistic symmetries.
Clearly the key challenge for such scenarios concerns the introduction of suitable mixing products, of which we provided several examples.
Our results preliminarily suggest that scenarios with non-universal deformation of relativistic symmetries
could in principle be realized in Nature and deserve dedicated experimental testing,
such as the mentioned studies looking for effects in the neutrino sector at a level of magnitude which is already
excluded for photons.
While we feel that such tests should not be postponed,
there are clearly still several tasks to be faced before fully establishing the consistency
of these scenarios. For example, it would be interesting to establish whether they 
are compatible with the setup of (possibly deformed) quantum field theories, though one should perhaps view this 
as a long-term goal, since even the understanding of quantum field theories with universal deformation of relativistic symmetries has still 
not reached full maturity.

Our mixing coproducts are in principle directly applicable to interactions between composite and fundamental particles 
(when governed by different relativistic properties). However, in some cases it should be possible to apply a constructive approach
for such mixing coproducts: ideally there might be cases in which one only needs to postulate
some universally-deformed relativistic properties for fundamental particles, then deriving the implied
(different) relativistic properties of various types of composite particles, and in such cases also the mixing coproducts
could be derived, rather then requiring a dedicated postulate. We feel that performing such derivations, even just in some 
particularly simple toy model, would be an important contributions to the further development
of the investigations we here reported.

For what concerns the broader picture of phenomenology it will be interesting to identify some characteristic observable differences
between the new scenarios of non-universal deformations of relativistic symmetries, on which we here focused, and the
scenarios in which relativistic symmetries are broken in particle-dependent manner, which have already been studied
for several years\cite{coleman1997cosmic,smeREVIEW}.

\section*{Acknowledgements}

We are grateful to Leonardo Barcaroli for contributing to the initial stages of this project.

\appendix

\section{Explicit derivation of mixing composition laws: solving the relevant differential equations}

In this Section we explain in some detail the procedure to obtain the deformation functions that appear in the composition laws for the mixing coproducts with trivial sum rule for energies, i.e. the deformation is present only in the spatial momentum sector. In doing so, we also show the general technique to solve all the differential equations contained in the main text. \\

The first set of equations we want to examine is given in Eq. \eqref{sistemaEp0,p1}. The operator for total infinitesimal boosts $N_{[p,k]}=hN_{[p]}+N_{[k]}$ acts on the total energy as

\begin{equation}
[N_{[p,k]},\varepsilon p_0+k_0]=h\Bigl\{\Bigr[{\partial \varepsilon \over \partial p_0}p_1+{\partial \varepsilon \over \partial p_1}\Bigl({1-e^{-2 \ell p_0} \over 2 \ell }-{\ell \over 2}p_1^2\Bigr)\Bigr]p_0+\varepsilon p_1\Bigr\}+k_1,
\end{equation}

thus defining the function $\varphi=\varepsilon p_0$ the first differential equation is

\begin{equation}
h\Bigl[{\partial \varphi \over \partial p_0}p_1+{\partial \varphi \over \partial p_1}\Bigl({1-e^{-2 \ell p_0} \over 2 \ell}-{\ell \over 2}p_1^2\Bigr)\Bigr]=fp_1,
\end{equation}

whose solution for $h$ is

\begin{equation}\label{dsd}h={fp_1 \over {\partial \varphi \over \partial p_0}p_1+{\partial \varphi \over \partial p_1}\Bigl({1-e^{-2\ell p_0} \over 2 \ell}-{\ell \over 2}p_1^2\Bigr)}.\end{equation}

At the same time we can compute the action of $N_{[p,k]}$ on the total spatial momentum i.e.

\begin{equation}
[N_{[p,k]},fp_1+k_1]=h\Bigl\{\Bigl[{\partial f \over \partial p_0}p_1+{\partial f \over \partial p_1}\Bigl({1-e^{-2 \ell p_0} \over 2 \ell}-{\ell \over 2}p_1^2\Bigr)\Bigr]p_1+f\Bigl({1-e^{-2 \ell p_0} \over 2 \ell}-{\ell \over 2}p_1^2\Bigr)\Bigr\}+k_0,
\end{equation}

that furnishes us a second equation\footnote{It is worth mentioning that, in this case, these two equations are not sufficient to find uniquely the three unknown functions $\varepsilon$, $f$ ed $h$.}

\begin{equation}\label{later}h\Bigl[{\partial \psi \over \partial p_0}p_1+{\partial \psi \over \partial p_1}\Bigl({1-e^{-2 \ell p_0} \over 2 \ell}-{\ell \over 2}p_1^2\Bigr)\Bigr]-\varphi=0,\end{equation}

with $\psi=fp_1.$ Now we can substitute $h$ from Eq. \eqref{dsd} and find

\begin{equation}
[{\partial \psi \over \partial p_0}p_1+{\partial \psi \over \partial p_1}\Bigl({1-e^{-2\ell p_0} \over 2 \ell}-{\ell \over 2}p_1^2\Bigr)\Bigr]\!-\!\varphi\Bigl[{\partial \varphi \over \partial p_0}p_1+{\partial \varphi \over \partial p_1}\Bigl({1-e^{-2 \ell p_0} \over 2 \ell}-{\ell \over 2}p_1^2\Bigr)\Bigr]\!=\!0.
\end{equation}

In order to simplify this equation we can define the function $\theta=\psi^2-\varphi^2$, which satisfies the differential equation

\begin{equation}\label{theta(p)}
{\partial \theta \over \partial p_0}+{1 \over p_1}{\partial \theta \over \partial p_1}\Bigl({1-e^{-2 \ell p_0} \over 2 \ell }-{\ell \over 2}p_1^2\Bigr)=0.
\end{equation}

The solution can be found by means of a pair of auxiliary variables $(\eta, \xi)$. For $\eta$ we choose

\begin{equation}\label{eta}
\eta=e^{\ell p_0}p_1^2-{4 \over \ell^2}\sinh^2\Bigl({\ell \over 2}p_0\Bigr)
\end{equation}

while $\xi$ must be a function of $p_0$ and $p_1$ such that it obeys

\begin{equation}
{\partial \over \partial \xi} \theta(\eta, \xi)={\partial \theta \over \partial p_0}{\partial p_0 \over \partial \xi}+{\partial \theta \over \partial p_1}{\partial p_1 \over \partial \xi}={\partial \theta \over \partial p_0}+{\partial \theta \over \partial p_1}{1 \over p_1}\Bigl({1-e^{-2 \ell p_0} \over 2 \ell}-{\ell \over 2}p_1^2\Bigr).
\end{equation}

On the light of this one can easily find that $\xi=p_0$ is a good choice and that Eq. \eqref{theta(p)} can be rewritten as follows

\begin{equation}
{\partial \theta \over \partial \xi}=0 \to \theta=C(\eta)=C\Bigl(e^{\ell p_0}p_1^2-{4 \over \ell^2}\sinh^2\Bigl({\ell \over 2}p_0\Bigr)\Bigr).
\end{equation}

Then, for instance, if we ask $f$ to be a function of $\varepsilon$ we have

\begin{equation}
f={\psi \over p_1}={1 \over |p_1|}\sqrt{\varphi^2+\theta^2}={1 \over |p_1|}\sqrt{\varepsilon^2p_0^2+C\Bigl(e^{\ell p_0}p_1^2-{4 \over \ell^2}\sinh^2\Bigl({\ell \over 2}p_0\Bigr)\Bigr)} .
\end{equation}

At this point we can use the fact that both functions $\varepsilon$ and $f$ must tend to $1$ when $l \to 0$. This implies that

\begin{equation}
C\Bigl(e^{\ell p_0}p_1^2-{4 \over \ell^2}\sinh^2\Bigl({\ell \over 2}p_0\Bigr)\Bigr)=e^{\ell p_0}p_1^2-{4 \over \ell^2}\sinh^2\Bigl({\ell \over 2}p_0\Bigr) .
\end{equation}

As a result, the final composition law is given by Eq. \eqref{box eps p0p1}.

The complementary case given by Eq. \eqref{box eps k0k1} can be dealt with in an analogous fashion, therefore we will not derive it explicitly here.

Instead we want to focus on the last system of equations in Eq. \eqref{sistema p0k0}.

From the first equation we deduce that $h=f(\partial \varphi / \partial k_0+1 )/ (\partial \varphi / \partial p_0)$ and $\varphi = \varepsilon p_0$. Collecting all the terms proportional to $p_1^2$ in the second equation we can rewrite the set as

\begin{equation}\label{sistema vinc}
\begin{cases}h\Bigl({\partial f \over \partial p_0}-{\ell \over 2}f\Bigr)=f{\partial f \over \partial k_0}\\ hf^2{1-e^{-2 \ell p_0} \over 2 \ell }=\varphi\end{cases},
\end{equation}

where we remind the reader that these equations must be satisfied under the validity of the constraint. Then, by substituting $h$ we have

\begin{equation}
{\partial f \over \partial p_0}-{{\partial \varphi \over \partial p_0} \over {\partial \varphi \over \partial k_0}+1}{\partial f \over \partial k_0}={\ell \over 2}f.
\end{equation}

Analogously to what we did to solve Eq. \eqref{p0k0} we define the new variables $\eta= (\varphi+k_0 )/ 2, \quad \xi= p_0-(\varphi+k_0)/2,$ such that
$p_0=\xi+\eta/2$ and thus $\partial p_0 / \partial \xi =1$. Deriving the equation for $\eta$ with respect to $\xi$, we obtain

\begin{align*}0&={1 \over 2}\Bigl({\partial \varphi \over \partial \xi}+{\partial k_0 \over \partial \xi}\Bigr)\\&={1 \over 2}\Bigl({\partial \varphi \over \partial p_0}{\partial p_0 \over \partial \xi}+{\partial \varphi \over \partial k_0}{\partial k_0 \over \partial \xi}+{\partial k_0 \over \partial \xi}\Bigr)\\&={1 \over 2}\Bigl[{\partial \varphi \over \partial p_0}+{\partial k_0 \over \partial \xi}\Bigl({\partial \varphi \over \partial k_0}+1\Bigr)\Bigr],\end{align*}

that implies

\begin{equation}
{\partial f(\xi,\eta) \over \partial \xi}={\partial f \over \partial p_0}{\partial p_0 \over \partial \xi}+{\partial f \over \partial k_0}{\partial k_0 \over \partial \xi}={\partial f \over \partial p_0}-{{\partial \varphi \over \partial p_0} \over {\partial \varphi \over \partial k_0}+1}{\partial f \over \partial k_0},
\end{equation}

and, thus, along the constraint $\varphi+k_0=0$, the equation $\partial f / \partial \xi=\ell f/ 2$. Its solution is $f=e^{\ell \xi/2}=e^{\ell p_0/2},$ i.e. the same we had for the analogous function in the system \eqref{(p+k)P p0}. At this point we can plug this solution in the second equation of Eq. \eqref{sistema vinc}, thereby finding

$${{\partial \varphi \over \partial p_0} \over {\partial \varphi \over \partial k_0}+1}\varphi={1 \over \ell }\sinh( \ell p_0).$$

In terms of the variables $\eta$ and $\xi$ we introduced above, this equation reads

\begin{equation}
\varphi{\partial \varphi \over \partial \xi}={1 \over \ell }\sinh( \ell \xi),
\end{equation}

which, given the usual limits for $\ell \to 0$, has solution $\varphi=2 \sinh(\ell \xi/2)/ \ell $.
In conclusion any pair of functions $(f,\varepsilon)$ that reduce to \eqref{(p+k)P p0} when $\varepsilon p_0+k_0=0$ provides a solution of this set of equations. \\

\section{Algebraic properties}
\label{sec:prop}

In this Section we analyze further some algebraic properties of the maps which have been introduced in the main text. In particular, we focus on the compatibility of the maps with both the algebra and coalgebra sectors. The relation with antipodes is also briefly discussed. Finally, we show how these maps can be used in order to naturally deduce the mixing coproducts. Again we concentrate on two specific examples which are particularly relevant for our intents, i.e. the maps relating the Poincar\'e with the $\kappa$-Poincar\'e algebra and vice-versa. The latter will be explicitly formulated in the bicrossproduct basis throughout this section. Moreover, in order to lighten the calculations but without any loss of generality, we will work in $1+1$ dimensions. Let us also remind that we call $( K_0, K_1, N_{[K]}) $ the generators of the Poincar\'e algebra and $(P_0, P_1, N_{[P]}) $ those of the $\kappa$-Poincar\'e algebra.

\subsection{Poincar\'e to $\kappa$-Poincar\'e maps}

We take into account the map, $\phi:\mathcal{P} \to \kappa\mathcal{P}$, whose explicit action on the generators of the Poincar\'e algebra is given by

\begin{equation}\label{PtokP}
\phi(K_0)={2 \over \ell}\sinh\Bigl({\ell \over 2}P_0\Bigr), \, \quad \phi(K_1)=e^{{\ell \over 2}P_0}P_1, \, \quad \phi(N_{[K]})={e^{{\ell \over 2}P_0} \over \cosh\Bigl({\ell \over 2}P_0\Bigr)}N_{[P]}.
\end{equation}

Let us start by discussing the compatibility of $\phi$ with the algebra and the antipodes and then we turn to coproducts. It is not difficult to verify that $\phi$ is compatible with both the commutators and the antipodes of $\mathcal{P}$. Indeed, for what regards the commutation relations of $\mathcal{P}$, we have

\begin{equation}
\phi([K_0,K_1])=\phi(0)=0=[\phi(K_0),\phi(K_1)],
\end{equation}

and also that

\begin{equation}
\phi([N_{[K]},K_0])=\phi(K_1)=e^{{\ell \over 2}P_0}P_1
\end{equation}

and

\begin{equation}
[\phi(N_{[P]}),\phi(K_0)]=\Biggl[{e^{{\ell \over 2}P_0} \over \cosh\Bigl({\ell \over 2}P_0\Bigr)}N_{[P]},{2 \over \ell}\sinh\Bigl({\ell \over 2}P_0\Bigr)\Biggr] ={e^{{ \ell \over 2}P_0} \over \cosh\Bigl({ \ell \over 2}P_0\Bigr)}\Bigl({d \over dP_0}{2 \over \ell}\sinh\Bigl({\ell \over 2}P_0\Bigr)\Bigr)P_1 =e^{{\ell \over 2}P_0}P_1.
\end{equation}

The same property also holds for the last commutator. In fact, it is easy to check that

\begin{equation}
\phi([N_{[K]},K_1])=\phi(K_0)={2 \over \ell}\sinh\Bigl({\ell \over 2}P_0\Bigr)
\end{equation}

and also

\begin{equation}
\begin{split}
[\phi(N_{[P]}),\phi(K_1)] =\Biggl[{e^{{\ell \over 2}P_0} \over \cosh\Bigl({\ell \over 2}P_0\Bigr)}N_{[P]},e^{{\ell \over 2}P_0}P_1\Biggr]={e^{{\ell \over 2}P_0} \over \cosh\Bigl({\ell \over 2}P_0\Bigr)}\Bigr[\cancel{{\ell \over 2}e^{{\ell \over 2}P_0}P_1^2}+e^{{\ell \over 2}P_0}\Bigl({1-e^{-2\ell P_0} \over 2 \ell}-\cancel{{\ell \over 2}P_1^2}\Bigr)\Bigr]\\={e^{\ell P_0} \over \cosh\Bigl({\ell \over 2}P_0\Bigr)}{1-e^{-2 \ell P_0} \over 2 \ell}={2 \over \ell}{{1 \over 2}\sinh(\ell P_0) \over \cosh\Bigl({\ell \over 2}P_0\Bigr)}={2 \over \ell}\sinh\Bigl({\ell \over 2}P_0\Bigr).
\end{split}
\end{equation}

As already anticipated, $\phi$ also respects the antipodal relations i.e.

\begin{align}&\phi(S(K_0))=\phi(-K_0)=-{2 \over \ell}\sinh\Bigl({\ell \over 2}P_0\Bigr)=S_{\kappa}\phi(K_0),\\
&\phi(S(K_1))=\phi(-K_1)=-e^{{\ell \over 2}P_0}P_1=e^{-{\ell \over 2}P_0}(-e^{\ell P_0}P_1)=S_{\kappa}\phi(K_1), \\
&\phi(S(N_{[K]}))=\phi(-N_{[K]})=-{e^{{\ell \over 2}P_0} \over \cosh\Bigl({\ell \over 2}P_0\Bigr)}N_{[P]}={e^{-{\ell \over 2}P_0} \over \cosh\Bigl({\ell \over 2}P_0\Bigr)}(-e^{\ell P_0}N_{[P]})=S_{\kappa}\phi(N_{[K]}) ,
\end{align}

where $S$ and $S_{\kappa}$ stand for the antipodes of the $\mathcal{P}$ and $\kappa\mathcal{P}$ algebras respectively.\\
Let us now look into what happens to the coalgebraic sector. As we can expect and have in fact highlighted in the main text, the map is not compatible with the coproducts. For instance, we find that

\begin{equation}
\phi\otimes\phi(\Delta K_0)=\phi\otimes\phi(K_0 \otimes\mathds{1}+\mathds{1}\otimes K_0)={2 \over \ell}\sinh\Bigl({\ell \over 2}P_0\Bigr)\otimes\mathds{1}+\mathds{1}\otimes{2 \over \ell}\sinh\Bigl({\ell \over 2}P_0\Bigr) ,
\end{equation}

while a different outcome is obtained with

\begin{equation}
\Delta(\phi(K_0)= {2 \over \ell}\Bigl({e^{{\ell \over 2}P_0}\otimes e^{{\ell \over 2}P_0}-e^{-{\ell \over 2}P_0}\otimes e^{-{\ell \over 2}P_0} \over 2}\Bigr)={2 \over\ell }\sinh\Bigl({\ell \over 2}P_0\Bigr)\otimes e^{{\ell \over 2}P_0}+e^{-{\ell \over 2}P_0}\otimes{2 \over\ell }\sinh\Bigl({\ell \over 2}P_0\Bigr),
\end{equation}

where we have used the formula $\Delta(e^{aP_0})=e^{aP_0}\otimes e^{aP_0}$ and performed some standard algebra manipulations. Despite there is no compatibility with the coproducts, it is interesting to notice that the following relation is realized

\begin{equation}
\Delta(\phi(K_1))=\Delta(e^{{\ell \over 2}P_0}P_1)=(e^{{\ell \over 2}P_0}\otimes e^{{\ell \over 2}P_0})(P_1\otimes\mathds{1}+e^{-lP_0}\otimes P_1)=e^{{\ell \over 2}P_0}P_1\otimes e^{{\ell \over 2}P_0}+e^{-{\ell \over 2}P_0}\otimes e^{{\ell \over 2}P_0}P_1,
\end{equation}

and thus

\begin{equation}
\Delta(\phi(K_{\mu}))=\phi(K_{\mu})\otimes e^{{\ell \over 2}P_0}+e^{-{\ell \over 2}P_0}\otimes\phi(K_{\mu}), \quad \mu=0,1.
\end{equation}

Finally, let us show how this map can be used to construct particle-dependent relativistic models. Specifically, $\phi$ is related to the mixing coproduct in Eq. \eqref{(p+k)P p0}. In fact, we can consider the composition, $\mathcal{P}\xrightarrow[]{\Delta}\mathcal{P}\otimes\mathcal{P}\xrightarrow[]{\phi\otimes\mathds{1}}\kappa\mathcal{P}\otimes\mathcal{P},$ that provides us with the mixing composition laws (or coproducts)

\begin{equation}
p\boxplus_{\mathcal{P}}k=\begin{cases}{2 \over \ell}\sinh\Bigl({\ell \over 2}p_0\Bigr)+k_0 \\ e^{{\ell \over 2}p_0}p_1+k_1\end{cases}.
\end{equation}

The main advantage of deriving the mixing coproducts in this way is that, by doing so, we can immediately identify whether the total momentum (resulted from the composition of a particle in Poincar\'e with another particle in $\kappa$-Poincar\'e ) is a charge associated to $\mathcal{P}$ or to $\mathcal{\kappa P}$. However, it is worth mentioning that there are some subtleties of these composition rules which derive from the non-compatibility of $\phi$ with the coproducts. In fact, this implies that for instance the two following ways of composing momenta, i.e.

\begin{equation}
\mathcal{P}\xrightarrow[]{\Delta}\mathcal{P}\otimes\mathcal{P}\xrightarrow[]{\phi\otimes\mathds{1}}\kappa\mathcal{P}\otimes\mathcal{P}\xrightarrow[]{\Delta\otimes \mathds{1}}\kappa\mathcal{P}\otimes\kappa\mathcal{P}\otimes\mathcal{P}
\end{equation}

and

\begin{equation}
\mathcal{P}\xrightarrow[]{\Delta}\mathcal{P}\otimes\mathcal{P}\xrightarrow[]{\Delta \otimes \mathds{1}}\mathcal{P}\otimes\mathcal{P}\otimes\mathcal{P}\xrightarrow[]{\phi\otimes\phi\otimes\mathds{1}}\kappa\mathcal{P}\otimes\kappa\mathcal{P}\otimes\mathcal{P}
\end{equation}

are different. As a result, if $p'\in\kappa\mathcal{P}$ and, thus,

$$(p\oplus_{\ell}p')\boxplus_{\mathcal{P}} k \ne p\boxplus_{\mathcal{P}}(p'\boxplus_{\mathcal{P}}k):$$

then this mixing coproduct is not associative with respect to the sum of $\kappa$-Poincar\'e\footnote{The reader can make a comparison with what we had for Eq. \eqref{isol-l'2}.}. In fact, we find explicitly

\begin{equation}
(p\oplus_{\ell}p')\boxplus_{\mathcal{P}} k=\begin{cases}{2 \over \ell}\sinh\Bigl({\ell \over 2}(p_0+p_0')\Bigr)+k_0 \\ e^{{\ell \over 2}(p_0+p_0')}(p_1+e^{-\ell p_0}p_1')+k_1\end{cases},
\end{equation}

that does not coincide with

\begin{equation}
p\boxplus_{\mathcal{P}}(p'\boxplus_{\mathcal{P}}k)=\begin{cases}{2 \over \ell}\sinh\Bigl({\ell \over 2}p_0\Bigr)+{2 \over \ell}\sinh\Bigl({\ell \over 2}p_0'\Bigr)+k_0\\e^{{\ell \over 2}p_0}p_1+e^{{\ell \over 2}p_0'}p_1'+k_1\end{cases}.
\end{equation}

On the other hand, it is evident that the associative property is still present if we consider the sum in Poincar\'e, i.e.

\begin{equation}
(p\boxplus_{\mathcal{P}} k)+k'=p\boxplus_{\mathcal{P}}(k+k'),
\end{equation}

where, according to our notation, $k'$ denotes the momentum of a particle obeying standard Poincar\'e symmetries. The opposite will happen for the mixing composition laws obtained from the inverse map, which we discuss below.

\subsection{ $\kappa$-Poincar\'e to Poincar\'e maps}

Now we wish to examine the properties of the inverse map $\tilde{\phi}:\kappa\mathcal{P}\to\mathcal{P}$, which is defined as

\begin{equation}\label{kPtoP}
\tilde{\phi}(P_0)={2 \over \ell} \ln\Bigl({\ell \over 2}K_0+\sqrt{{\ell ^2 \over 4}K_0^2+1}\Bigr), \, \quad \tilde{\phi}(P_1)={K_1 \over {\ell \over 2}K_0+\sqrt{{\ell ^2 \over 4}K_0^2+1}},\, \quad \tilde{\phi}(N_{[P]})={\sqrt{{\ell ^2 \over 4}K_0^2+1} \over {\ell \over 2}K_0+\sqrt{{\ell ^2 \over 4}K_0^2+1}}N_{[K]}
\end{equation}

and, thanks to linearity, can be extended to the whole $\kappa\mathcal{P}$. First of all, let us verify that $\tilde{\phi}$ is actually the inverse of $\phi$. This follows from the following formulas:

\begin{equation}
\phi \circ \tilde{\phi}(P_0)={2 \over \ell}\ln\Bigl[\sinh\Bigl({\ell \over 2}P_0\Bigr)+\sqrt{\sinh^2\Bigl({\ell \over 2}P_0\Bigr)+1}\Bigr]=P_0
\end{equation}

and also

\begin{equation}
\phi \circ \tilde{\phi}(P_1)={e^{{\ell \over 2}P_0}P_1 \over \sinh\Bigl({\ell \over 2}P_0\Bigr)+\sqrt{\sinh^2\Bigl({\ell \over 2}P_0\Bigr)+1}}={e^{{\ell \over 2}P_0}P_1 \over \sinh\Bigl({\ell \over 2}P_0\Bigr)+\cosh\Bigl({\ell \over 2}P_0\Bigr)}=P_1,
\end{equation}

while

\begin{equation}
\phi \circ \tilde{\phi}(N_{[P]})={\sqrt{\sinh^2\Bigl({\ell \over 2}P_0\Bigr)+1} \over \sinh\Bigl({\ell \over 2}P_0\Bigr)+\sqrt{\sinh^2\Bigl({\ell \over 2}P_0\Bigr)+1}}{e^{{\ell \over 2}P_0} \over \cosh\Bigl({\ell \over 2}P_0\Bigr)}N_{[K]}=N_{[P]}.
\end{equation}

To conclude the demonstration that $\tilde{\phi} = \phi^{-1}$ we have also that

\begin{align*}
&\tilde{\phi}\circ\phi(K_0)={2 \over \ell} \sinh\Bigl[\ln\Bigl({\ell \over 2}K_0+\sqrt{{\ell ^2 \over 4}K_0^2+1}\Bigr)\Bigr]=K_0,\\
&\tilde{\phi}\circ\phi(K_1)=e^{\ln\Bigl({\ell \over 2}K_0+\sqrt{{\ell ^2 \over 4}K_0^2+1}\Bigr)}{K_1 \over {\ell \over 2}K_0+\sqrt{{\ell ^2 \over 4}K_0^2+1}}=K_1,\\
&\tilde{\phi}\circ\phi(N_{[K]})={\sqrt{{\ell ^2 \over 4}K_0^2+1} \over {\ell \over 2}K_0+\sqrt{{\ell ^2 \over 4}K_0^2+1}}{e^{\ln\Bigl({\ell \over 2}K_0+\sqrt{{\ell ^2 \over 4}K_0^2+1}\Bigr)} \over \cosh\Bigl[\ln\Bigl({\ell \over 2}K_0+\sqrt{{\ell ^2 \over 4}K_0^2+1}\Bigr)\Bigr]}N_{[P]}=N_{[K]}.
\end{align*}

Analogously to what we did in the previous subsection, let us start looking at the behavior of the $\kappa$-Poincar\'e algebra under this map. Again we find that both commutators and antipodes are compatible with the map. Indeed, we have

\begin{equation}
\tilde{\phi}([P_0,P_1])=\tilde{\phi}(0)=0=[\tilde{\phi}(P_0),\tilde{\phi}(P_1)]
\end{equation}

and also

\begin{equation}
\tilde{\phi}([N_{[P]},P_0])=\tilde{\phi}(P_1)={K_1 \over {\ell \over 2}K_0+\sqrt{{\ell^2 \over 4}K_0^2+1}},
\end{equation}

and the same result follows from

\begin{equation}
[\tilde{\phi}(N_{[P]}),\tilde{\phi}(P_0)]=\Biggl[{\sqrt{{\ell^2 \over 4}K_0^2+1} \over {\ell \over 2}K_0+\sqrt{{\ell^2 \over 4}K_0^2+1}}N_{[K]},{2 \over \ell}\ln\Bigl({\ell \over 2}K_0+\sqrt{{\ell^2 \over 4}K_0^2+1}\Bigr)\Biggr]={K_1 \over {\ell \over 2}K_0+\sqrt{{\ell^2 \over 4}K_0^2+1}}.
\end{equation}

To complete the proof of the compatibility with the algebra, it remains to be checked the commutator $\tilde{\phi}([N_{[P]},P_1])$. It is still straightforward but more involved to obtain that

\begin{equation}
\tilde{\phi}([N_{[P]},P_1])=\tilde{\phi}\Bigl({1-e^{-2lP_0} \over 2\ell}-{\ell \over 2}P_1^2\Bigr)=K_0{\sqrt{{\ell^2 \over 4}K_0^2+1} \over \Bigl({\ell \over 2}K_0+\sqrt{{\ell^2 \over 4}K_0^2+1}\Bigr)^2}-{\ell \over 2}\Biggl({K_1 \over {\ell \over 2}K_0+\sqrt{{\ell^2 \over 4}K_0^2+1}}\Biggr)^2,
\end{equation}

and again

\begin{equation}
[\tilde{\phi}(N_{[P]}),\tilde{\phi}(P_1)]=K_0{\sqrt{{\ell^2 \over 4}K_0^2+1} \over \Bigl({\ell \over 2}K_0+\sqrt{{\ell^2 \over 4}K_0^2+1}\Bigr)^2}-{\ell \over 2}\Biggl({K_1 \over {\ell \over 2}K_0+\sqrt{{\ell^2 \over 4}K_0^2+1}}\Biggr)^2 .
\end{equation}

These three relations ensure the compatibility of the map with the $\kappa$-Poincar\'e algebra. \\
Now we can turn to the compatibility with the antipodes. It is rather easy to show that $\tilde{\phi}$ is compatible with the antipodes. Following the procedure above, the reader can check that that

\begin{align}
&\tilde{\phi}(S_{\kappa}(P_0))=\tilde{\phi}(-P_0)=-{2 \over \ell}\ln\Bigl({\ell \over 2}K_0+\sqrt{{\ell^2 \over 4}K_0^2+1}\Bigr)=S(\tilde{\phi}(P_0)),\\&\tilde{\phi}(S_{\kappa}(P_1))=\tilde{\phi}(-e^{\ell P_0}P_1)=-\Bigl({\ell \over 2}K_0+\sqrt{{\ell^2 \over 4}K_0^2+1}\Bigr)^2{K_1 \over {\ell \over 2}K_0+\sqrt{{\ell^2 \over 4}K_0^2+1}}=S(\tilde{\phi}(P_1))
\end{align}

and finally

\begin{equation}
\tilde{\phi}(S_{\kappa}(N_{[P]})) =\tilde{\phi}(-e^{\ell P_0}N_{[P]}) =-{\Bigl(-{\ell \over 2}K_0+\sqrt{{\ell^2 \over 4}K_0^2+1}\Bigr)\sqrt{{\ell^2 \over 4}K_0^2+1} \over \Bigl(-{\ell \over 2}K_0+\sqrt{{\ell^2 \over 4}K_0^2+1}\Bigr)^2}N_{[K]} =S(\tilde{\phi}(N_{[P]})).
\end{equation}

In the light of the above discussion, again we do not expect to have compatibility with the coalgebra. For the sake of brevity, we do not explicitly show that $\tilde{\phi}$ is not compatible with the coproducts. However, let us just mention that a shortcut is provided by realizing that the non-compatibility is a direct consequence of the fact that

\begin{equation}
\ln(x+y+\sqrt{(x+y)^2+1})\ne \ln(x+\sqrt{x^2+1})+\ln(y+\sqrt{y^2+1}).
\end{equation}

Just as we did in the previous subsection, for the purposes of this work it is finally interesting to see how the composition laws for particles obeying different symmetry structures can be built from $\tilde{\phi}$. It is noteworthy to say that, being the coproduct of the $\kappa$-Poincar\'e algebra non-commutative, we need to distinguish two possible cases, i.e. $\kappa\mathcal{P}\xrightarrow[]{\Delta}\kappa\mathcal{P}\otimes\kappa\mathcal{P}\xrightarrow[]{\tilde{\phi}\otimes\mathds{1}}\mathcal{P}\otimes\kappa\mathcal{P}$ and $\kappa\mathcal{P}\xrightarrow[]{\Delta}\kappa\mathcal{P}\otimes\kappa\mathcal{P}\xrightarrow[]{\mathds{1}\otimes\tilde{\phi}}\kappa\mathcal{P}\otimes\mathcal{P}$. These two chains of maps give rise to the mixing coproducts (which we have already introduced)

\begin{align}\label{(p+k)kP k0}
\begin{split}
&p\boxplus_{\kappa\mathcal{P}}k=\begin{cases}p_0+{2 \over \ell} \ln\Bigl({\ell \over 2}k_0+\sqrt{{\ell ^2 \over 4}k_0^2+1}\Bigr)\\p_1+e^{-lp_0}{k_1 \over {\ell \over 2}k_0+\sqrt{{\ell ^2 \over 4}k_0^2+1}}\end{cases},\\
&k\boxplus_{\kappa\mathcal{P}}p=\begin{cases}{2 \over \ell} \ln\Bigl({\ell \over 2}k_0+\sqrt{{\ell ^2 \over 4}k_0^2+1}\Bigr)+p_0\\{k_1 \over {\ell \over 2}k_0+\sqrt{{\ell ^2 \over 4}k_0^2+1}}+{1 \over \Bigl({\ell \over 2}k_0+\sqrt{{\ell ^2 \over 4}k_0^2+1}\Bigr)^2}p_1\end{cases}.
\end{split}
\end{align}

As one may expect, now the behavior of these composition laws under associativity is inverted with respect to the mixing coproducts obtained with $\phi$. In fact, these composition laws enjoy associativity (from both left and right) with respect to the sum of $\kappa$-Poincaré, i.e.

\begin{align}
&k\boxplus_{\kappa\mathcal{P}}(p\oplus_{\ell}p')=(k\boxplus_{\kappa\mathcal{P}}p)\oplus_{\ell}p',\\
&(p'\oplus_{\ell}p)\boxplus_{\kappa\mathcal{P}}k=p'\oplus_{\ell}(p\boxplus_{\kappa\mathcal{P}}k),
\end{align}

as the reader can straightforwardly verify. On the other hand, these ways of composing momenta are not associative if we consider the usual sum in Poincar\'e. This is a direct consequence of the fact that we do not have compatibility with the coproducts. In fact, it is easy to realize that

\begin{equation}
p\boxplus_{\kappa\mathcal{P}}(k+k')=
\begin{cases}
p_0+{2 \over \ell} \ln\Bigl[{ \ell \over 2}(k_0+k_0')+\sqrt{{ \ell ^2 \over 4}(k_0+k_0')^2+1}\Bigr]\\p_1+e^{-lp_0}{k_1+k_1' \over { \ell \over 2}(k_0+k_0')+\sqrt{{ \ell ^2 \over 4}(k_0+k_0')^2+1}}
\end{cases},
\end{equation}

while a different result is obtained by the expression

\begin{equation}
(p\boxplus_{\kappa\mathcal{P}}k)\boxplus_{\mathcal{P}}k'\!=\!\begin{cases}p_0\!+\!{2 \over \ell } \ln\Bigl({ \ell \over 2}k_0+\sqrt{{ \ell ^2 \over 4}k_0^2+1}\Bigr)\!+\!{2 \over \ell } \ln\Bigl({ \ell \over 2}k_0'+\sqrt{{ \ell ^2 \over 4}k_0'^2+1}\Bigr)\\p_1\!+\!e^{-lp_0}{k_1 \over { \ell \over 2}k_0+\sqrt{{ \ell ^2 \over 4}k_0^2+1}}\!+\!{e^{-lp_0} \over \Bigl({ \ell \over 2}k_0+\sqrt{{ \ell ^2 \over 4}k_0^2+1}\Bigr)^2}{k_1' \over { \ell \over 2}k_0'+\sqrt{{ \ell ^2 \over 4}k_0'^2+1}}\end{cases}.
\end{equation}

\section{(2+1)-dimensional case}
\label{sec:2D}

Here we wish to understand if the results of this work, which have been explicitly carried out only in $(1+1)$ dimensions in the main text, can be extended to the $(2+1)$ case. This would also represent an important preliminary test-bed for the four-dimensional case of physical interest. Of course, the main difference (and potential source of complexity) with respect to the $(1+1)$ dimensional case is that now we also have to deal with the rotation generator $R_{[K]}$ and $R_{[P]}$ \footnote{Note that we only have one generator of rotations since there are two spatial directions.}, respectively for the Poincar\'{e} and the $k$-Poincar\'{e} algebra \footnote{As usually done in this paper we denote with $( K_0, K_1, R_{[K]}, N_{[K]}) $ the generators of the Poincar\'e algebra and with $(P_0, P_1, R_{[P]}, N_{[P]}) $ those of the $\kappa$-Poincar\'e algebra.}.

We shall see here that imposing the covariance under rotational symmetry gives rather tight constraints on the form of the deformation functions in the mixing coproducts.

Let us start introducing the linear map $\phi:\mathcal{P}\to\kappa\mathcal{P}$ given by

\begin{align}\label{rot?}\begin{split}
&\phi(K_0)=\varepsilon(P_0,P_1,P_2)P_0,\\
&\phi(K_i)=f_{i}(P_0,P_1,P_2)P_{i},\\
&\phi(N_{[K]_i})=h_{i}(P_0,P_1,P_2)N_{[P]_i}+\sigma_{i}(P_0,P_1,P_2)R_{[P]},\\
&\phi(R_{[K]})=\omega(P_0,P_1,P_2)R_{[P]},
\end{split}\end{align}

where $i=1,2$. In order to find the unknown functions, we ask $\phi$ to be compatible with the commutation relations of the symmetry algebras. This will allow us to fully determine and solve the problem, up to a single degree of freedom.

It is trivial to verify that

$$\phi([K_{\mu},K_{\nu}])=\phi(0)=0, \quad \mu,\nu=0,1,2$$
$$[\phi(K_{\mu}),\phi(K_{\nu})]=0,$$

while instead the condition

$$[\phi(R_{[K]}),\phi(K_0)]=\phi([R_{[K]},K_0])=0$$

gives us

\begin{equation}\label{cond1}
\omega[R_{[P]},\varepsilon P_0]=0.
\end{equation}

From Eq. \eqref{cond1} we get the following differential equation

$$\omega\Bigl({\partial \varepsilon \over \partial P_1}P_2-{\partial \varepsilon \over \partial P_2}P_1\Bigr)P_0=0, $$

which means that

\begin{equation}\label{eps(P0,modP)}
\varepsilon=\varepsilon(P_0,P)
\end{equation}

with $P \equiv | \vec{P} | $.

Analogously from imposing the equality between

$$\phi([R_{[K]},K_i])=\phi(\epsilon_{ij}K_j)=\epsilon_{ij}f_{j}P_j, \quad i,j=1,2$$

and

$$[\phi(R_{[K]}),\phi(K_i)]=\omega[R_{[P]},f_{i}P_i],$$

we get

\begin{equation}\label{cond2}
\omega\Bigl({\partial f_i \over \partial P_1}P_2-{\partial f_i \over \partial P_2}P_1\Bigr)P_i+\epsilon_{ij}(\omega f_{i}-f_{j})P_j=0.
\end{equation}

From this last equation one can easily obtain the following conditions

\begin{align}
\begin{split}\label{f(P0modP)}
&\omega=1, \\
& f_{i}\equiv f=f(P_0,P).
\end{split}
\end{align}

Acting in a similar way with the commutator $ [R_{[K]},N_{[K]_i}] $, we get

\begin{equation}\label{h(P0modP)}
h_{i}\equiv h=h(P_0,P)
\end{equation}

and the system of differential equations

\begin{equation}\label{cond3}
\begin{cases}{\partial \sigma_1 \over \partial P_1}P_2-{\partial \sigma_1 \over \partial P_2}P_1=\sigma_2 \\ {\partial \sigma_2 \over \partial P_1}P_2-{\partial \sigma_2 \over \partial P_2}P_1=-\sigma_1\end{cases}.
\end{equation}

It is convenient to express the system \eqref{cond3} in polar coordinates

$$\begin{cases}-{\partial \sigma_1 \over \partial \theta}=\sigma_2 \\ {\partial \sigma_2 \over \partial \theta}=\sigma_1\end{cases},$$

where $P_1 \equiv \rho\cos\theta$ and $ P_2 \equiv \rho\sin\theta$.

Thus the functions $\sigma_i$ must obey the harmonic oscillator equation, i.e.,

$$-\sigma_2={\partial \sigma_1 \over \partial \theta}=-A(\rho)\sin\theta+B(\rho)\cos\theta,$$

from which we get in cartesian coordinates

\begin{align*}
&\sigma_1={A(P) \over P}P_1+{B(P) \over P}P_2,\\
&\sigma_2=-{B(P) \over P}P_1+{A(P) \over P}P_2.
\end{align*}

At this point we are only left with imposing covariance under the action of the boosts, which, as usual, involves setting the equality between

$$\phi([N_{[K]_1},K_0])=\phi(K_1)=fP_1,$$

and

\begin{equation}\label{cond inter}
[\phi(N_{[K]_1}),\phi(K_0)]=\Bigl[hN_{[P]_1}+\Bigl({A \over P}P_1+{B \over P}P_2\Bigr)R_{[P]},\varepsilon P_0\Bigr].
\end{equation}

We first note that

$$[N_{[P]_i},\varepsilon]={\partial \varepsilon \over \partial P_0}[N_{[P]_i},P_0]+{\partial \varepsilon \over \partial P}[N_{[P]_i},P],$$

where

\begin{align*}
[N_{[P]_1},P] = [N_{[P]_1},\sqrt{P_1^2+P_2^2}] =
{P_1 \over P}\Bigl({1-e^{-2 \ell P_0} \over 2 \ell}-{\ell \over 2}P^2\Bigr)
\end{align*}

and

$$[N_{[P]_2},P]={P_2 \over P}\Bigl({1-e^{-2 \ell P_0} \over 2 \ell}-{\ell \over 2}P^2\Bigr).$$

It is trivial to see that the rotational generator gives a null contribution to Eq. \eqref{cond inter}, thus we get

\begin{equation}\label{cond4}
h\Bigl\{\Bigl[{\partial \varepsilon \over \partial P_0}P_i+{\partial \varepsilon \over \partial P}{P_i \over P}\Bigl({1-e^{-2 \ell P_0} \over 2 \ell}-{\ell \over 2}P^2\Bigr)\Bigr]P_0+\varepsilon P_i\Bigr\}=fP_i.
\end{equation}

Defining as usual $\varphi \equiv \varepsilon P_0 $ we get

\begin{equation}\label{h}
h={f \over {\partial \varphi \over \partial P_0}+{1 \over P}\Bigl({1-e^{-2 \ell P_0} \over 2 \ell}-{\ell \over 2}P^2\Bigr){\partial \varphi \over \partial P}}.
\end{equation}

Now imposing the equality between

$$[\phi(N_{[K]_1}),\phi(K_2)]=\Bigl[hN_{[P]_1}+\Bigl({A \over P}P_1+{B \over P}P_2\Bigr)R_{[P]},fP_2\Bigr]$$

and

$$\phi([N_{[K]_1},K_2])=0,$$

we get the following differential equation

$$h\Bigl[{\partial f \over \partial P_0}+{\partial f \over \partial P}{1 \over P}\Bigl({1-e^{-2 \ell P_0} \over 2 \ell}-{\ell \over 2}P^2\Bigr)- \ell f\Bigr]P_2-f\Bigl({A \over P}P_1+{B \over P}P_2\Bigr)=0.$$

Since $f = f(P_0, P)$ we get $A=0$ and

\begin{equation}\label{B}
B={h \over f}\Bigl[{\partial f \over \partial P_0}+{\partial f \over \partial P}{1 \over P}\Bigl({1-e^{-2 \ell P_0} \over 2 \ell}-{\ell \over 2}P^2\Bigr)- \ell f\Bigr]P.
\end{equation}

Now imposing $[\phi(N_{[K]_1}),\phi(K_1)]$ be equal to

$$\phi([N_{[K]_1},K_1])=\phi(K_0)=\varphi,$$

we get

\begin{align}
\begin{split}\label{cond5}
&h\Bigl\{\Bigl[{\partial f \over \partial P_0}P_1+{\partial f \over \partial P}{P_1 \over P}\Bigl({1-e^{-2 \ell P_0} \over 2 \ell}-{\ell \over 2}P^2\Bigr)\Bigr]P_1+f\Bigl({1-e^{-2lP_0} \over 2 \ell}+{\ell \over 2}P^2-{\ell \over 2}P_1^2\Bigr)\Bigr\}+f{B \over P}P_2^2=\varphi.
\end{split}
\end{align}

Using Eq. \eqref{B} we can rewrite Eq. \eqref{cond5} as

\begin{equation}\label{condend}
h\Bigl\{\Bigl[{\partial f \over \partial P_0}P+{\partial f \over \partial P}\Bigl({1-e^{-2 \ell P_0} \over 2 \ell}-{\ell \over 2}P^2\Bigr)\Bigr]P+f\Bigl({1-e^{-2 \ell P_0} \over 2 \ell}-{ \ell \over 2}P^2\Bigr)\Bigr\}=\varphi.
\end{equation}

Given the form of the function $h$ in Eq. \eqref{h}, Eq. \eqref{condend} is identical to Eq. \eqref{later} up to the usual definition $\psi=fP$.

This means that the functions $h$ and $f$ are the same found in Sec. \ref{sec:classificazione}, where one has to substitute $P$ with $| \vec{P} | $, i.e.,

\begin{align}\label{eps2D}
\begin{split}&\varepsilon=\varepsilon(P_0,P),\\&f={1 \over P}\sqrt{\varepsilon^2P_0^2-{4 \over \ell^2}\sinh^2\Bigl({\ell \over 2}P_0\Bigr)+e^{ \ell P_0}P^2},\\&h={\sqrt{\varepsilon^2P_0^2-{4 \over \ell^2}\sinh^2\Bigl({\ell \over 2}P_0\Bigr)+e^{\ell P_0}P^2} \over P{\partial \varphi \over \partial P_0}+\Bigl({1-e^{-2 \ell P_0} \over 2 \ell}-{\ell \over 2}P^2\Bigr){\partial \varphi \over \partial P}}.\end{split}
\end{align}

From Eq. \eqref{eps2D} and Eq. \eqref{B} one can easily obtain

\begin{equation}\label{B'}
B={\varphi-hf\Bigl({1-e^{-2 \ell P_0} \over 2 \ell }-{ \ell \over 2}P^2\Bigr) \over P f}.
\end{equation}

Finally in the explicit case in which $\varepsilon={2 \over \ell P_0}\sinh\Bigl({\ell \over 2}P_0\Bigr)$, we get that the map $\phi$ is given by

\begin{align}\label{phi2D}
\begin{split}
&\phi(K_0)={2 \over \ell}\sinh\Bigl({\ell \over 2}P_0\Bigr),\\
&\phi(K_i)=e^{{\ell \over 2}P_0}P_i,\\
&\phi(R_{[K]})=R_{[P]},\\
&\phi(N_{[K]_i})={e^{{\ell \over 2}P_0} \over \cosh\Bigl({\ell \over 2}P_0\Bigr)}(N_{[P]_i}+{\ell \over 2}\epsilon_{ij}P_j R_{[P]}).
\end{split}
\end{align}

The inverse map

$$\tilde{\phi}:\kappa\mathcal{P}\to\mathcal{P},$$

whose derivation is left to the reader, is given by

\begin{align}\label{tildephi2D}
\begin{split}
&\tilde{\phi}(P_0)={2 \over \ell}\ln\Bigl({\ell \over 2}K_0+\sqrt{{\ell^2 \over 4}K_0^2+1}\Bigr),\\
&\tilde{\phi}(P_i)={K_i \over {\ell \over 2}K_0+\sqrt{{\ell^2 \over 4}K_0^2+1}},\\
&\tilde{\phi}(R_{[P]})=R_{[K]},\\
&\tilde{\phi}(N_{[P]_i})={\sqrt{{\ell^2 \over 4}K_0^2+1} \over {\ell \over 2}K_0+\sqrt{{\ell^2 \over 4}K_0^2+1}}\Biggl(N_{[K]_i}-{\ell \over 2}\epsilon_{ij}{K_i \over \sqrt{{\ell^2 \over 4}K_0^2+1}}R_{[K]}\Biggr).
\end{split}
\end{align}


\begin{thebibliography}{80}
\addcontentsline{toc}{chapter}{Bibliografia}

\bibitem{amelino2013quantum} G. Amelino-Camelia, \textit{Quantum spacetime phenomenology}, Living Rev.Rel. 16 (2013) 5
\bibitem{magueijo2003generalized} J. Maguejo, L. Smolin, \textit{Generalized Lorentz invariance with an invariant energy scale}, Phys. Rev. D 67, 044017 (2003)
\bibitem{amelino2012fate} G. Amelino-Camelia, \textit{Fate of Lorentz symmetry in relative-locality momentum spaces}, Phys. Rev. D 85, 084034 (2012)
\bibitem{amelino2001testable} G. Amelino-Camelia, \textit{Testable scenario for Relativity with minimum-length}, Phys.Lett. B510 (2001) 255-263
\bibitem{amelino2011principle} G. Amelino-Camelia, L. Freidel, J. Kowalski-Glikman, L. Smolin, \textit{The principle of relative locality}, Phys. Rev. D 84, 084010 (2011)
\bibitem{amelinoDSRijmpd1135} G. Amelino-Camelia, \textit{Doubly Special Relativity}, Nature 418:34-35, 2002
\bibitem{amelino2010doubly} G. Amelino-Camelia, \textit{Doubly-Special Relativity: Facts, Myths and Some Key Open Issues}, Symmetry 2010, 2, 230-271
\bibitem{kowalski2003non} J. Kowalski-Glikman, \textit{Doubly Special Relativity and quantum gravity phenomenology}, \href{https://arxiv.org/abs/hep-th/0312140}{arXiv:hep-th/0312140v1} [gr-qc] (2003)
\bibitem{freidelDESITTERSNYDER} A. Banburski, L. Friedel, \textit{Snyder momentum space in relative locality}, Phys. Rev. D 90, 076010 (2014)
\bibitem{balena} G. Amelino-Camelia, G. Palmisano, G. Gubitosi, \textit{Pathways to relativistic curved momentum spaces: de Sitter case study}, International Journal of Modern Physics DVol. 25, No. 02, 1650027 (2016)
\bibitem{agostini2004hopf} G. Amelino-Camelia, A. Agostini, F. D'Andrea, \textit{Hopf-algebra description of noncommutative space-time symmetries}, International Journal of Modern Physics AVol. 19, No. 30, pp. 5187-5219 (2004)
\bibitem{gacmixingold} G. Amelino-Camelia, \textit{Particle-Dependent Deformations of Lorentz Symmetry}, Symmetry 2012, 4(3), 344-378
\bibitem{coleman1997cosmic} S. Coleman, S. L. Glashow, \textit{Cosmic ray and neutrino tests of special relativity}, Physics Letters B Volume 405, Issues 3–4, 24 July 1997, Pages 249-252
\bibitem{smeREVIEW} D. Colladay, A. Kostelecky, \textit{Lorentz-Violating Extension of the Standard Model}, Phys.Rev.D58:116002, 1998
\bibitem{grbgac1998} G. Amelino-Camelia, J. Ellis, N.E. Mavromatos, D.V. Nanopoulos, S. Sarkar, \textit{Potential Sensitivity of Gamma-Ray Burster Observations to Wave Dispersion in Vacuo}, Nature 393:763-765,1998
\bibitem{gacSMOLINprd2009} G. Amelino-Camelia, L. Smolin, \textit{Prospects for constraining quantum gravity dispersion with near term observations}, Phys.Rev.D80:084017, 2009
\bibitem{amelino2015icecube} G. Amelino-Camelia, D. Guetta, T. Piran, \textit{IceCube neutrinos and Lorentz invariance violation}, Astrophys.J. 806 (2015) no.2, 269
\bibitem{jacob2007neutrinos} U. Jacob, T. Piran, \textit{GRBs Neutrinos as a Tool to Explore Quantum Gravity induced Lorentz Violation}, NaturePhys.3:87-90, 2007
\bibitem{Amelino-Camelia:2016ohi} G. Amelino-Camelia, L. Barcaroli, G. D'Amico, N. Loret, G. Rosati \textit{IceCube and GRB neutrinos propagating in quantum spacetime}, Phys.Lett. B761 (2016) 318-325
\bibitem{amelino2016icecube} Giovanni Amelino-Camelia, Leonardo Barcaroli, Giacomo D'Amico, Niccoló Loret, Giacomo Rosati, \textit{Quantum-gravity-induced dual lensing and IceCube neutrinos}, Int.J.Mod.Phys. D26 (2017) 1750076
\bibitem{paperbyMA} M. G. Aartsen et al., \textit{Neutrino Interferometry for High-Precision Tests of Lorentz Symmetry with IceCube}, Nature Physicsvolume 14, pages961–966 (2018)
\bibitem{abdo2009limit} A. A. Abdo et al., \textit{A limit on the variation of the speed of light arising from quantum gravity effects}, Nature. 2009 Nov 19;462(7271):331-4
\bibitem{aharonian2008limits} F. Aharonian et al., \textit{Upper Limits from HESS AGN Observations in 2005-2007}, Astronomy\&Astrophysics 478, 387-393 (2008)
\bibitem{amelino2011relative} G. Amelino-Camelia, L. Freidel, J. Kowalski-Glikman, L. Smolin, \textit{Relative locality and the soccer ball problem}, Phys. Rev. D 84, 087702 (2011)
\bibitem{barcaroli2014relative} L. Barcaroli, \textit{Relative Locality framework, with and without gravity}, Ph. D. thesis (2014)
\bibitem{kosinski1994classical} P. Kosinski, J. Lukierski, P. Maslanka, J. Sobczyk, \textit{The classical basis for $\kappa$-deformed Poincaré (super)algebra and the second $\kappa$-deformed supersymmetric Casimir}, Mod. Phys. Lett. A10 (1995) 2599
\bibitem{Borowiec} A. Borowiec, A. Pacol, \textit{Classical basis for kappa-Poincare algebra and doubly special relativity theories}, J.Phys.A43:045203, 2010




\end{thebibliography}
\end{document}